\documentclass[fleqn,usenatbib]{mnras}

\usepackage{newtxtext,newtxmath}

\usepackage[T1]{fontenc}

\DeclareRobustCommand{\VAN}[3]{#2}
\let\VANthebibliography\thebibliography
\def\thebibliography{\DeclareRobustCommand{\VAN}[3]{##3}\VANthebibliography}

\usepackage{xcolor,xspace}
\usepackage{threeparttable}
\usepackage{orcidlink}
\usepackage{graphicx}	
\usepackage{amsmath}	

\newcommand{\nustar}{{\it NuSTAR}\xspace}
\newcommand{\swift}{{\it Swift}\xspace}
\newcommand{\nicer}{{\it NICER}\xspace}
\newcommand{\fermi}{{\it Fermi}\xspace}
\newcommand{\LEIA}{{\it LEIA}\xspace}
\newcommand{\EP}{{\it EP}\xspace}

\newcommand{\xspec}{\texttt{XSPEC}\xspace}
\newcommand{\addspec}{\texttt{addspec}\xspace}
\newcommand{\grppha}{\texttt{grppha}\xspace}
\newcommand{\wxtpip}{\texttt{wxtpipeline}\xspace}
\newcommand{\wxtpro}{\texttt{wxtproducts}\xspace}
\newcommand{\xrtpip}{\texttt{xrtpipeline}\xspace}
\newcommand{\xsel}{\texttt{xselect}\xspace}
\newcommand{\ximage}{\texttt{ximage}\xspace}
\newcommand{\nupip}{\texttt{nupipeline}\xspace}
\newcommand{\nupro}{\texttt{nuproducts}\xspace}
\newcommand{\barycorr}{\texttt{barycorr}\xspace}
\newcommand{\efsearch}{\texttt{efsearch}\xspace}

\newcommand{\uergcms}{erg cm$^{-2}$ s$^{-1}$}
\newcommand{\chsq}{$\chi^{2}$}

\title[RX J0520.5-6932: the 2024 outburst]{Broadband study of the Be X-ray binary RX J0520.5-6932 during its outburst in 2024}

\author[H. N. Yang et al.]{
H. N. Yang\orcidlink{0000-0002-7680-2056}$^{1,2,3}$\thanks{E-mail: hnyang@nao.cas.cn},
C. Maitra\orcidlink{0000-0002-0766-7313}$^{2}$,
G. Vasilopoulos\orcidlink{0000-0003-3902-3915}$^{4,5}$\thanks{E-mail: gevas@phys.uoa.gr},
F. Haberl\orcidlink{0000-0002-0107-5237}$^{2}$,
P. A. Jenke$^{6}$,
A. S. Karaferias$^{7}$,  \newauthor
R. Sharma\orcidlink{0000-0003-0366-047X}$^{8}$,
A. Beri\orcidlink{0000-0003-3753-3102}$^{9,10}$,
L. Ji\orcidlink{0000-0001-9599-7285}$^{11}$,
C. Jin\orcidlink{0000-0002-2006-1615}$^{1,3}$,
W. Yuan$^{1,3}$, 
Y. J. Zhang$^{12}$,   
C. Y. Wang$^{12}$,   
X. P. Xu$^{1,3}$, \newauthor
Y. Liu$^{1}$,
W. D. Zhang$^{1}$,
C. Zhang$^{1,3}$,
Z. X. Ling$^{1,3}$,
H. Y. Liu$^{1}$, 
H. Q. Cheng\orcidlink{0000-0003-4200-9954}$^{1}$,
and H. W. Pan$^{1}$
\\
$^{1}$National Astronomical Observatories, Chinese Academy of Sciences, 20A Datun Road, Beijing 100101, China\\
$^{2}$Max-Planck-Institut f\"{u}r extraterrestrische Physik, Gie\ss{}enbachstra\ss{}e 1, D-85748 Garching bei M\"{u}nchen, Germany\\
$^{3}$School of Astronomy and Space Science, University of Chinese Academy of Sciences, 19A Yuquan Road, Beijing 100049, China\\
$^{4}$Department of Physics, National and Kapodistrian University of Athens, University Campus Zografos, GR 15784, Athens, Greece\\
$^{5}$Institute of Accelerating Systems \& Applications, University Campus Zografos, Athens, Greece\\
$^{6}$University of Alabama in Huntsville, Huntsville, AL 35805, USA\\
$^{7}$Ludwig-Maximilians-Universität München, 80333 München, Germany\\ 
$^{8}$Raman Research Institute, C.V. Raman Avenue, Sadashivanagar, Bengaluru 560080, Karnataka, India\\
$^{9}$Indian Institute of Science Education and Research (IISER) Mohali, Punjab 140306, India \\
$^{10}$Physics \& Astronomy, University of Southampton, Southampton, Hampshire SO17 1BJ, UK\\
$^{11}$School of Physics and Astronomy, Sun Yat-sen University, Zhuhai, 519082, People's Republic of China\\
$^{12}$Department of Astronomy, Tsinghua University, Beĳing 100084, China\\
}

\date{Accepted XXX. Received YYY; in original form ZZZ}

\pubyear{2024}

\begin{document}
\label{firstpage}
\pagerange{\pageref{firstpage}--\pageref{lastpage}}
\maketitle

\begin{abstract}
A new giant outburst of the Be X-ray binary RX J0520.5-6932 was detected and subsequently observed with several space-borne and ground-based instruments. This study presents a comprehensive analysis of the optical and X-ray data, focusing on the spectral and timing characteristics of selected X-ray observations. 
A joint fit of spectra from simultaneous observations performed by the X-ray telescope (XRT) on the Neil Gehrels Swift Observatory (\swift) and Nuclear Spectroscopic Telescope ARray (\nustar) provides broadband parameter constraints, including a cyclotron resonant scattering feature (CRSF) at $32.2_{-0.7}^{+0.8}$ keV with no significant energy change since 2014, and a weaker Fe line. 
Independent spectral analyses of observations by the Lobster Eye Imager for Astronomy (\LEIA), Einstein Probe(\EP), \swift-XRT, and \nustar demonstrate the consistency of parameters across different bands.
Luminosity variations during the current outburst were tracked. The light curve of the Optical Gravitational Lensing Experiment (OGLE) aligns with the X-ray data in both 2014 and 2024.
Spin evolution over 10 years is studied after adding \fermi Gamma-ray Burst Monitor (GBM) data, improving the orbital parameters, with an estimated orbital period of 24.39 days, slightly differing from OGLE data.
Despite intrinsic spin-up during outbursts, a spin-down of $\sim$0.04s over 10.3 years is suggested. 
For the new outburst, the pulse profiles indicate a complicated energy-dependent shape, with decreases around 15 keV and 25 keV in the pulsed fraction, a first for an extragalactic source.
Phase-resolved \nustar data indicate variations in parameters such as flux, photon index, and CRSF energy with rotation phase.

\end{abstract}

\begin{keywords}
pulsars: individual: RX J0520.5-6932-- Magellanic Clouds -- X-rays: binaries
\end{keywords}



\section{Introduction}
Be/X-ray binaries (BeXRBs) are a subclass of high-mass X-ray binaries (HMXBs) that consists of a Be star and a compact object, typically a neutron star (NS) (see \citealt{Reig2011} for a review). 
Most BeXRBs have a transient nature, characterized by two types of outbursts. Type I outbursts are periodic and occur at the periastron passage of the NS, originating from the interaction with the Be star's circumstellar disk. Type II outbursts are more intense and less frequent, often associated with significant changes in the Be star's disk.

RX J0520.5-6932 (hereafter J0520) is located in the Large Magellanic Cloud (LMC) and was initially discovered by ROSAT observations \citep{Schmidtke1994}, while it has historically exhibited both Type I and major outbursts. 
A major outburst was reported to happen in 1995 which was detected in optical and X-ray \citep{Edge2004}.
During a Type I outburst, J0520 showed coherent X-ray pulsations at $\sim$8.04 s and spectral properties consistent with a BeXRB \citep{Vasilopoulos2014}.
In 2014, its X-ray luminosity reached levels near the Eddington limit for a NS \citep{Vasilopoulos2014ATel}. During this period, observations with Nuclear Spectroscopic Telescope ARray (\nustar; \citealt{Harrison2013}) revealed a cyclotron resonant scattering feature (CRSF), indicating a strong magnetic field \citep{Tendulkar2014}. Recent studies have modelled the accretion torque and orbital parameters of this system, providing insights into the behaviour of matter under extreme conditions \citep{Karaferias2023}.

At the end of March 2024, a new outburst from J0520 was detected \citep{Semena2024ATel, Sharma2024ATel, Zhang2024ATel} by a number of instruments, including the Mikhail Pavlinsky Astronomical Roentgen Telescope - X-ray Concentrator (ART-XC) telescope on board the Spektr-Roentgen-Gamma ({\it SRG}) observatory \citep{Pavlinsky2021}, the Neutron star Interior Composition Explorer (\nicer) \citep{Gendreau2016}, the Wide-field X-ray Telescope (WXT) on board the newly launched Einstein Probe mission (\EP; \citealt{Yuan2022}), as well as the \EP pathfinder Lobster Eye Imager for Astronomy (\LEIA; \citealt{ZhangChen_2022, Ling2023}).
A \nustar Directors Discretionary Time (DDT) observation (PI: C. Maitra) was performed to characterize the hard X-ray spectrum. The Neil Gehrels Swift Observatory (\swift; \citealt{Gehrels_2004}) followed the evolution of the soft X-ray emission with several monitoring observations, with one such observation performed within the \nustar observation period, facilitating a joint analysis. Additionally, long-term monitoring of LMC by \LEIA provided the variation on the soft X-ray flux and parameters of the object. \nicer also performed high-cadence monitoring throughout the outburst phase.
Further analysis was also performed with data from the Optical Gravitational Lensing Experiment (OGLE; \citealt{Udalski1992}) and \fermi Gamma-ray Burst Monitor (GBM; \citealt{Meegan2009}).

This paper is organized as follows. We describe the observations and data reduction in Section~\ref{sec2}. Then in Section~\ref{sec3} we detail the X-ray properties derived from the various datasets, including spectral and timing analyses. In Section~\ref{sec4}, we discuss the characteristics of this outburst and compare it with the 2014 outburst. Section~\ref{sec5} summarizes the results and conclusions of this work.

\section{Observations and Data Reduction}\label{sec2}

\subsection{Multiwavelength Data}

The earliest report on the current outburst from J0520 dates back to UT 2024 March 29 by ART-XC on board the {\it SRG} observatory \citep{Semena2024ATel}, which was promptly confirmed by \nicer and \LEIA \citep{Sharma2024ATel, Zhang2024ATel}.
In fact, an increase in the X-ray flux of this source has already been detected by \EP-WXT as early as March 22 during its commissioning phase.
Since its launch in 2022, the \EP pathfinder \LEIA has been monitoring the LMC on a daily basis. These data, either in single snapshots (several hundred seconds) or after stacking, reveal no significant X-rays from the source prior to the \EP-WXT detection on the 22nd. 
Therefore, the outburst is likely to have just started around March 22 and peaked around March 29. A \nustar DDT observation was promptly performed following the reported detection, which was jointed by a simultaneous \swift observation. Subsequently, \swift carried out multiple observations of the object.
The outburst of J0520 in 2014 was also observed by \nustar, offering the opportunity to analyze its X-ray behaviour over a long timescale of a decade.
Detailed observational logs are presented in Table~\ref{tab:obs_log}, including data from two \nustar observations 80001002002 and 80001002004 in 2014, hereafter referred to as observations 2014n1 and 2014n2, respectively.
Moreover, \nicer also performed high-cadence follow-up observations since April, which we include in this work for flux comparison and the detailed study will be presented in Sharma et al. (in preparation).

\begin{figure}
    \includegraphics[width=\columnwidth]{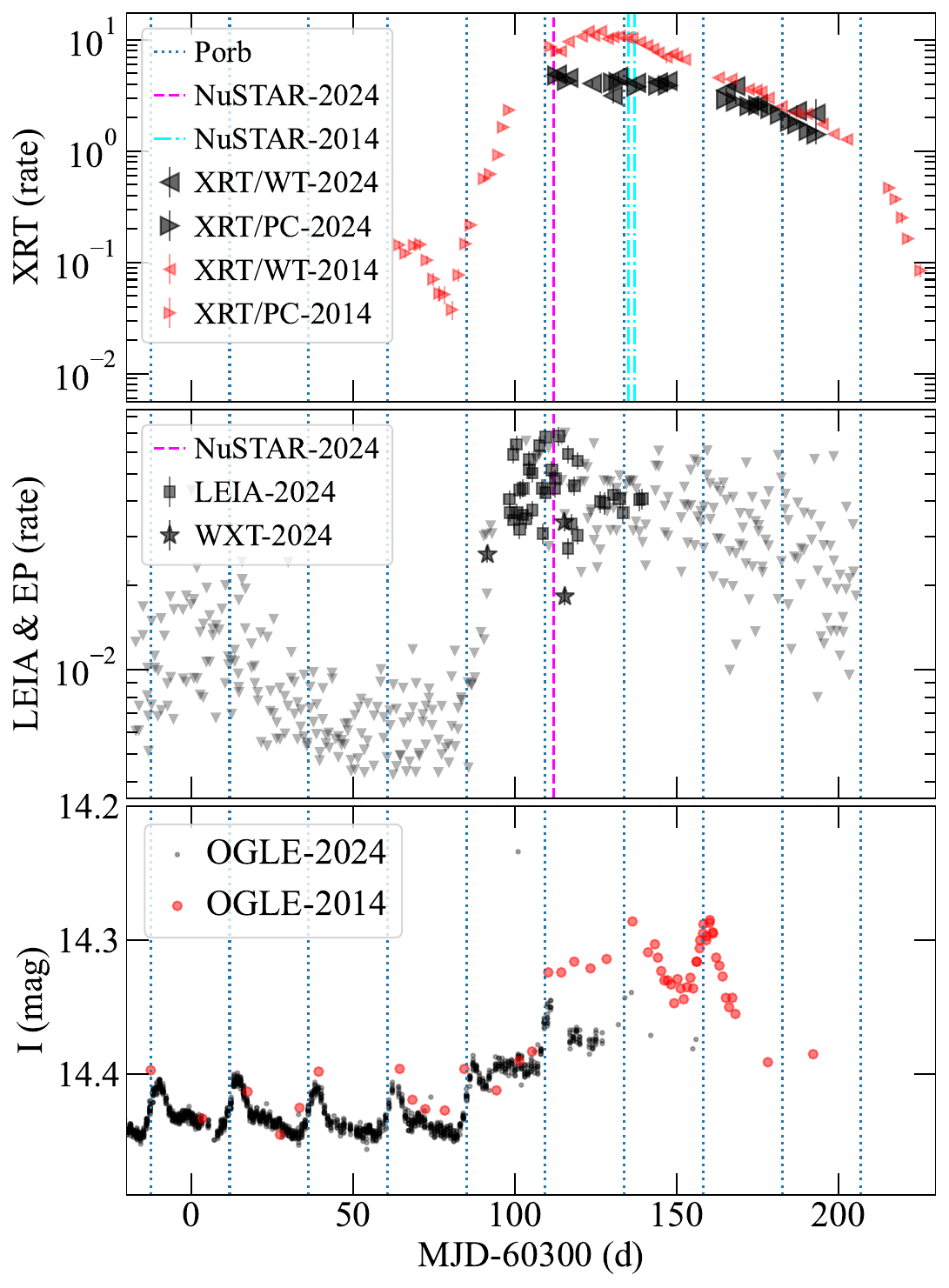}
    \caption{The optical and X-ray light curves of the 2024 outburst. Within the same panels, the data from the 2014 outburst were overplotted for visual comparison, which (marked as red) are shifted by an integer number (i.e. 154) of $P_{\rm orb}=24.39~d$, a value derived by the updated ephemeris (see Section~\ref{sec:new_orb}). Then the 0-epoch of 2014 data corresponds to MJD-56544~d.
    Vertical dotted lines mark orbital cycles phased at the periastron passages based on the parameters in Section~\ref{sec:new_orb}.
    Top: \swift-XRT light curves in 2024 (black triangle) and 2014 (red triangle), with the magenta and cyan dashed lines representing the time of \nustar observations in 2024 and 2014, respectively.
    Middle panel: \LEIA (black square) and \EP-WXT (black star) light curves in 2024, with upper limits plotted as grey triangles if the object was not detected in one-shot observations. Rebinning can provide tighter flux constraints, which will be presented later. The magenta dashed line represents the time of \nustar observations in 2024, the same as the middle panel.
    Bottom panel: OGLE light curves in 2024 (black dot) and 2014 (red dot).}
    \label{fig:multiband}
\end{figure}

In addition to X-ray observations, the brightening of J0520 in optical was also recorded previously(e.g. \citealt{Edge2004}).
As a long-term optical survey project, OGLE 
started monitoring the LMC during the second phase of the survey \citep{Udalski1997}. J0520 is listed in the X-Ray variables OGLE Monitoring (XROM) system\footnote{\url{https://ogle.astrouw.edu.pl/ogle4/xrom/xrom.html}} \citep{Udalski2008} and monitored continuously to this day. For purposes of comparison, we plotted the OGLE I band magnitudes for the counterpart of J0520 and the \swift-XRT count rates over time in both 2024 and 2014 outbursts in Fig.~\ref{fig:multiband}. The light curves of \LEIA and \EP-WXT during the current outburst are also plotted in the same figure.

\subsection{\LEIA and \EP}
J0520 was detected by \LEIA on March 29, 2024 \citep{Zhang2024ATel}, and has been monitored by \LEIA 1-2 times daily since the initial detection. 
As the pathfinder of \EP-WXT, the data of \LEIA were processed and calibrated using the standard data analysis tool \wxtpip (version 0.1.0) and the calibration database (CALDB) designed for the Einstein Probe (Liu et al. in prep), to produce cleaned event files \citep{ZhangChen_2022}. The CALDB is generated based on the results of on-ground and in-orbit calibration campaigns of the LEIA instrument (\citealt{2024ChengEA} and Cheng et al. in prep).
Products including spectrum and light curve were extracted using the \wxtpro tool in \wxtpip, during which a circular region with a radius of 67 pixels (1 pixel $\simeq$ 0.136 arcmin) and an annulus region with inner and outer radii of 134 and 268 pixels were defined as the source and background regions, respectively.

To improve spectral fitting, the spectra collected during various snapshots (based on the count rate and observation cadence) were combined using the \addspec tool assuming exposure as weights (it provided consistent results when counts rather than exposure was used as weights), which ensures a minimum of 200 net counts in each combined spectrum.

In fact, the LMC was observed by \EP-WXT on March 22 and April 15, during which the source was detected. This pushed the estimated beginning epoch of the current outburst back by approximately one week. The data were processed in the same way as for \LEIA by using the \EP-WXT CALDB instead.

\begin{figure*}
    \includegraphics[width=2\columnwidth]{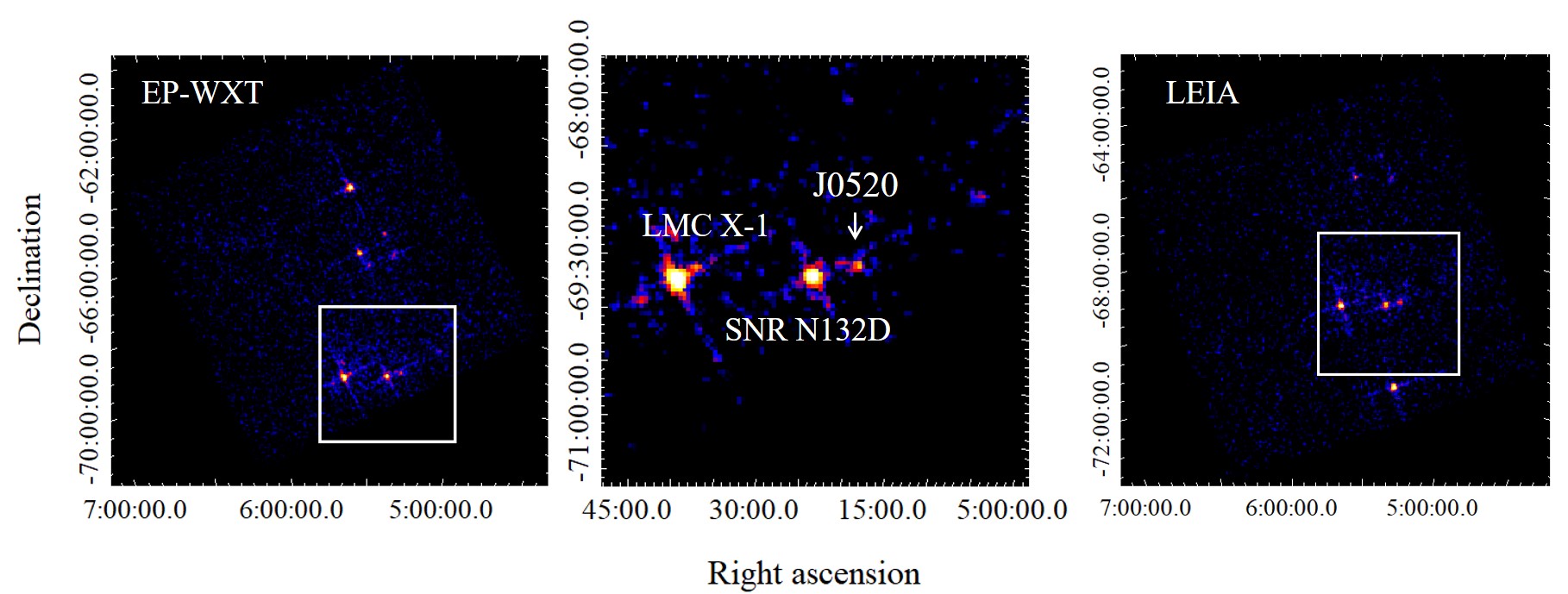}
    \caption{
    The single-exposure images from \EP-WXT and \LEIA with exposure times of 1210s and 840s, respectively. Left panel: 1-CMOS \EP-WXT image ($9.3^{\circ} \times 9.3^{\circ}$), covering 1/48 of the total \EP-WXT field of view. Middle panel: In the \EP-WXT image, a $4^{\circ}\times 4^{\circ}$ region around J0520 is enlarged, with nearby bright sources labeled. Right panel: 1-CMOS \LEIA image ($9.3^{\circ} \times 9.3^{\circ}$), covering 1/4 of the total \LEIA field of view. The $4^{\circ}\times 4^{\circ}$ region of the middle panel is marked with a white box in the left and right panels. 
    }
    \label{fig:img_ep_leia}
\end{figure*}

\subsection{\swift}
During the new outburst, J0520 was observed several times by \swift from April to July 2024 (see Table~\ref{tab:obs_log}). 
The standard \xrtpip version 0.13.5 and CALDB version 20240506 were used for the \swift-XRT data reduction of these observations. After extracting the images with the \xsel tool, we used \ximage to determine the pile-up effect on PC mode data.
For PC mode, the source region was an annulus with an inner radius of 6 pixels and an outer radius of 20 pixels to reduce the pile-up effect, where 1 pixel corresponds to 2.357 arcsec. 
For the background region, a circle with a radius of 50 pixels was used.
For WT mode, source and background photons were extracted from circular regions with radii of 8 pixels.

\subsection{\nustar}
J0520 was observed by \nustar on UT 2024 April 12 for an exposure time of 18.9 ks.
The \nustar data were reduced with \nupip version 0.4.8 and CALDB version 20240315. Circular regions with radii of 100 arcsec were defined as source and background regions, respectively. Spectra and light curves of FPMA and FPMB instruments were then generated with \nupro version 0.3.2. Photon times were corrected to the equivalent time at the solar system barycenter by using the \barycorr tool.

\subsection{\fermi}
The GBM Accreting Pulsars Program (GAPP) has been particularly successful in monitoring X-ray pulsars during outbursts and providing light-curves and spin period evolution measurements via an online database\footnote{GBM Accreting Pulsars Program: \url{https://gammaray.msfc.nasa.gov/gbm/science/pulsars.html}} \citep{2020ApJ...896...90M}. For each source pulse frequency and pulsed flux measurements in the 12-50 keV band using the NaI detectors are provided. We used publicly available GBM data for our modelling.

\section{X-ray Properties}\label{sec3}

\subsection{Spectral Analysis}
\label{sec:spec} 

\subsubsection{Joint Fit} \label{sec:joint_fit}

All the spectra extracted from \LEIA, \EP-WXT, \swift-XRT and \nustar observations were fitted using \xspec version 12.11.1 \citep{1996ASPC..101...17A}. 
Initially, we performed a joint fit of the \swift-XRT and \nustar spectra using the data from April 12 and 13, which were rebinned using \grppha to ensure at least 1 count per bin. 
C-statistic was used to evaluate the fit. 
Abundances from \citet{Wilms2000} and cross-sections from \citet{Verner1996} were adopted. The cross-normalization constant was frozen at unity for XRT (PC mode) and allowed to vary for XRT (WT mode), FPMA, and FPMB. Uncertainties hereafter indicate 90\% confidence intervals unless stated otherwise.

During outbursts, many BeXRBs exhibit a X-ray spectral shape that can be described by a power-law model with an exponential cutoff \citep[e.g.][]{Coburn2002}. Preliminary fits indicate that the power-law model combined with a high-energy cutoff (\texttt{highecut}) and cutoff power-law (\texttt{cutoffpl}) deviates from the data at high and low energies, respectively.
We fit the joint spectra using a power-law model with a Fermi-Dirac cutoff (\texttt{constant*tbabs*powerlaw*fdcut}) as the continuum. However, the simple continuum fitting did not match the data satisfactorily, with a cstat value of 2951.79 for 2796 degrees of freedom (dof). By adding an absorption component to account for a CRSF, the fitting significantly improves to cstat/dof=2744.17/2793. There is also a Fe emission line component present at $\sim$6.5 keV. By including a Gaussian line the cstat/dof for the joint fit could further be reduced to 2692.40/2790. The equivalent width of the Fe line is $46^{+18}_{-13}$ eV. The specific spectral parameters are listed in Table~\ref{tab:fit_simul1}.

\begin{figure}
    \includegraphics[width=\columnwidth]{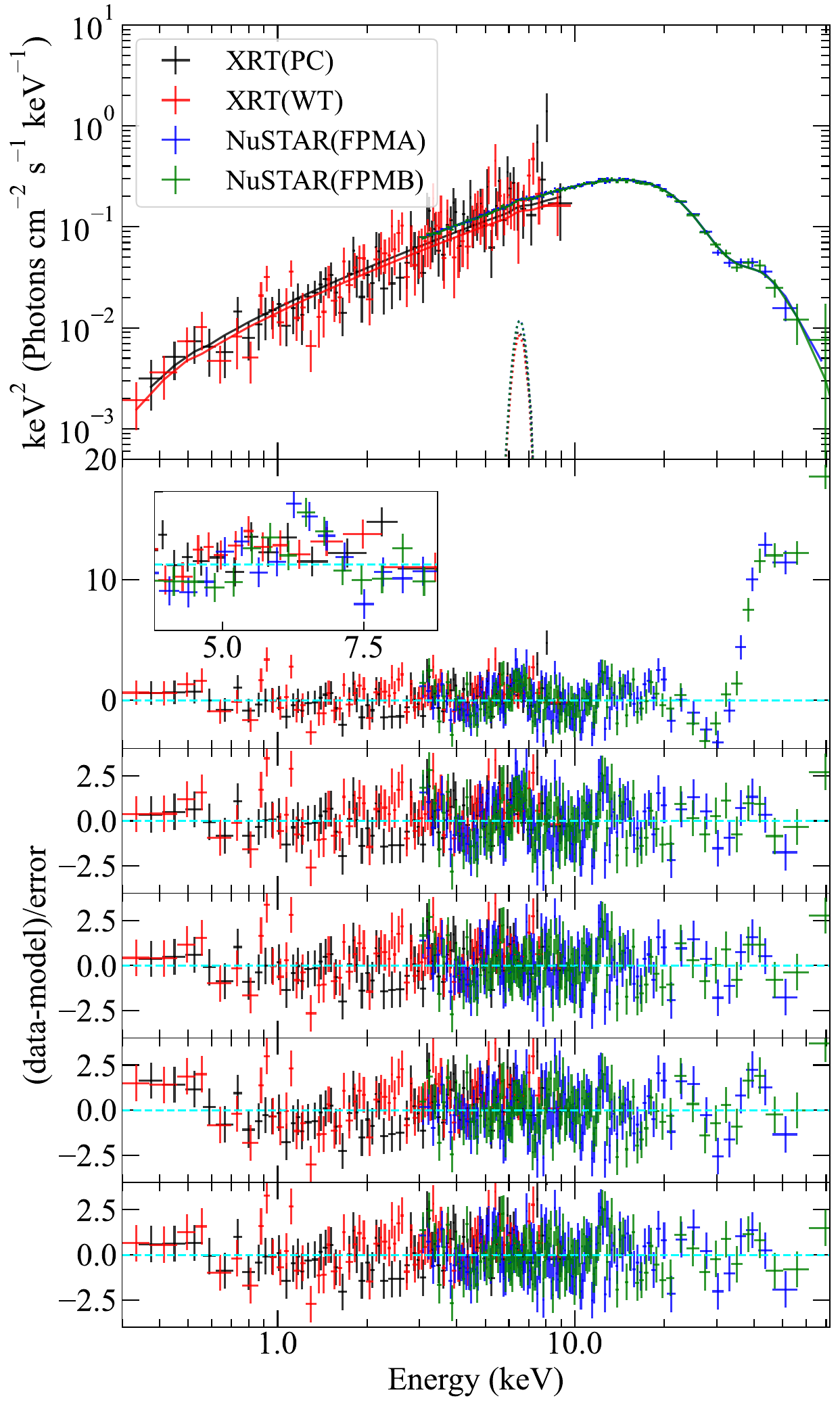}
    \caption{\swift-XRT (black for PC and red for WT) and \nustar (blue for FPMA and green for FPMB) unfolded spectra with fit results and residuals for different models. Top panel: spectra with their best-fit \texttt{const*tbabs(powerlaw*fdcut*gabs+gauss)} models. 
    Second panel: residual of the continuum-only model: \texttt{const*tbabs*powerlaw*fdcut}. 
    Third panel: residual of the continuum model after adding an absorption component indicating a CRSF: \texttt{const*tbabs*powerlaw*fdcut*gabs}. 
    Fourth panel: residual of the best-fit model: \texttt{const*tbabs(powerlaw*fdcut*gabs+gauss)}. 
    Fifth panel: residual of the model: \texttt{const*tbabs(nthcomp*gabs+gauss)}. 
    Bottom panel: residual of the model: \texttt{const*tbabs(cutoffpl*gabs+gauss+bbody)}
    }
    \label{fig:fit_simul}
\end{figure}

Additionally, both of the other two models used by \citet{Tendulkar2014}, a thermally Comptonized continuum (\texttt{nthcomp}) and a cutoff power-law (\texttt{cutoffpl}) with a blackbody component (\texttt{bbody}), provided a worse fit for the data, whose parameters are listed in Table~\ref{tab:fit_simul2}. 
The data with the best-fit model and residuals of different models are shown in Fig.~\ref{fig:fit_simul}. 
Under different models, the CRSF energy is fitted consistently, indicating that this key parameter is model-independent. 
We also considered a partial covering model, which has been used to describe HMXBs (e.g. \citealt{Furst2014, Xiao2024}), with a simple cutoff power-law. However, this model did not provide an improved fit.
We consider \texttt{const*tbabs(powerlaw*fdcut*gabs+gauss)} as the best-fit model for the simultaneous observations of \swift-XRT and \nustar, which also enables a more effective comparison with previous results in 2014.

\begin{table}
\caption{Joint fit results.}
\label{tab:fit_simul1}
\begin{threeparttable}
\begin{tabular}{llr}
\hline
\multicolumn{1}{c}{Component} & \multicolumn{1}{c}{Parameter} & \multicolumn{1}{c}{Value}            \\ \hline
\multicolumn{3}{l}{\texttt{const*tbabs(powerlaw*fdcut)}}                                                   \\
\texttt{constant} $^{\mathrm{a}}$ & $C_{\mathrm{WT}}$ & $0.893_{-0.111}^{+0.124}$ \\
 & $C_{\mathrm{FPMA}}$ & $1.170_{-0.101}^{+0.113}$ \\
 & $C_{\mathrm{FPMB}}$ & $1.162_{-0.100}^{+0.113}$ \\
\texttt{tbabs} & $N_\mathrm{H}$ & $0.065_{-0.044}^{+0.053}$ \\
\texttt{powerlaw} & PhoIndex & $0.793_{-0.022}^{+0.021}$ \\
 & norm & $0.019_{-0.002}^{+0.002}$ \\
\texttt{fdcut} & $E_{\mathrm{cutoff}}$ (keV) & $14.9_{-0.4}^{+0.4}$ \\
 & $E_{\mathrm{fold}}$ (keV) & $5.28_{-0.10}^{+0.10}$ \\
total & cstat/dof & 2951.79/2796 \\
\multicolumn{3}{l}{\texttt{const*tbabs(powerlaw*fdcut*gabs)}}                                              \\
\texttt{constant} $^{\mathrm{a}}$ & $C_{\mathrm{WT}}$ & $0.893_{-0.111}^{+0.124}$ \\
 & $C_{\mathrm{FPMA}}$ & $1.147_{-0.099}^{+0.110}$ \\
 & $C_{\mathrm{FPMB}}$ & $1.140_{-0.098}^{+0.110}$ \\
\texttt{tbabs} & $N_\mathrm{H}$ & $0.036_{-0.036}^{+0.051}$ \\
\texttt{powerlaw} & PhoIndex & $0.687_{-0.041}^{+0.039}$ \\
 & norm & $0.021_{-0.002}^{+0.002}$ \\
\texttt{fdcut} & $E_{\mathrm{cutoff}}$ (keV) & $10.3_{-1.6}^{+1.5}$ \\
 & $E_{\mathrm{fold}}$ (keV) & $7.58_{-0.31}^{+0.32}$ \\
\texttt{gabs} & $E_{\mathrm{CRSF}}$ (keV) & $32.1_{-0.7}^{+0.7}$ \\
 & $\sigma_{\mathrm{CRSF}}$ (keV) & $6.48_{-0.56}^{+0.63}$ \\
 & Strength & $14.5_{-2.5}^{+3.0}$ \\
total & cstat/dof & 2744.17/2793 \\
\multicolumn{3}{l}{\texttt{const*tbabs(powerlaw*fdcut*gabs+gauss)}}                                        \\
\texttt{constant} $^{\mathrm{a}}$ & $C_{\mathrm{WT}}$ & $0.893_{-0.112}^{+0.124}$ \\
 & $C_{\mathrm{FPMA}}$ & $1.168_{-0.101}^{+0.113}$ \\
 & $C_{\mathrm{FPMB}}$ & $1.161_{-0.100}^{+0.112}$ \\
\texttt{tbabs} & $N_\mathrm{H}$ & $0.046_{-0.042}^{+0.051}$ \\
\texttt{powerlaw} & PhoIndex & $0.750_{-0.044}^{+0.046}$ \\
 & norm & $0.021_{-0.002}^{+0.002}$ \\
\texttt{fdcut} & $E_{\mathrm{cutoff}}$ (keV) & $12.8_{-1.8}^{+2.1}$ \\
 & $E_{\mathrm{fold}}$ (keV) & $7.49_{-0.32}^{+0.33}$ \\
\texttt{gabs} & $E_{\mathrm{CRSF}}$ (keV) & $32.2_{-0.7}^{+0.8}$ \\
 & $\sigma_{\mathrm{CRSF}}$ (keV) & $7.00_{-0.66}^{+0.80}$ \\
 & Strength & $16.4_{-3.1}^{+4.1}$ \\
\texttt{gauss} & $E_{\mathrm{Fe}}$ (keV) & $6.51_{-0.10}^{+0.09}$ \\
 & $\sigma_{\mathrm{Fe}}$ (keV) & $0.273_{-0.104}^{+0.147}$ \\
 & norm & $0.000164_{-0.000046}^{+0.000058}$ \\
total & cstat/dof & 2692.40/2790 \\ \hline

\end{tabular}
\begin{tablenotes}
    \item[a] cross-normalization constants 
\end{tablenotes}

\end{threeparttable}
\end{table}

\subsubsection{Fit for Individual Observations} \label{sec:individual_fit}

Next, the \LEIA and \swift-XRT data from individual epochs were fitted by an absorbed power-law model with hydrogen column density $N_{\mathrm{H}}$ frozen at $4.6\times 10^{20}$ cm$^{-2}$, the value of the best-fit model from the joint fit. 
The same value was also used as $N_{\mathrm{H}}$ for the individual \nustar observation. 
Only the \nustar observation, which could be rebinned to achieve at least 20 photons in each bin, is suitable to use \chsq statistic. The FPMA and FPMB spectra were tied during the fit, with the constant parameter for FPMA frozen as unity but FPMB free. 
As for the other models used in Section \ref{sec:joint_fit}, while a fit with a blackbody component also yields satisfactory \chsq statistics, it does not provide physically reasonable parameters. 
In the case of \swift-XRT and \LEIA, there is also at least 1 count per bin, and C-statistic is used to estimate the goodness of fit due to the insufficient statistical quality \citep{Kaastra_2017}.

\begin{figure}
    \includegraphics[width=\columnwidth]{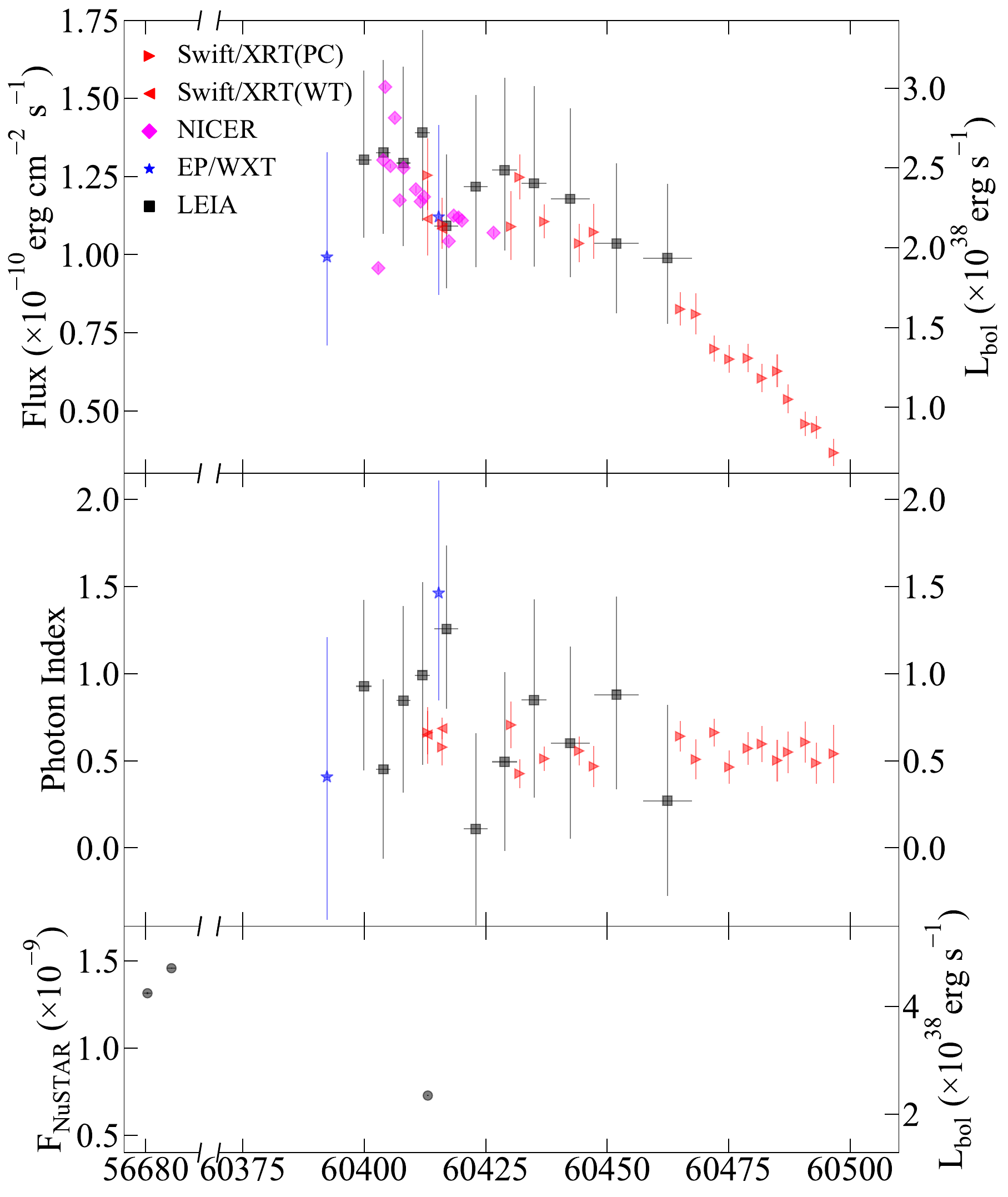}
    \caption{\LEIA and \swift-XRT light curves. Top panel: the evolution of unabsorbed flux in 1.5-5 keV from \LEIA and \swift-XRT and corresponding bolometric X-ray luminosity using the correction based on the joint spectral fit. 
    For comparison, the \nicer data of J0520 during the same period are also included, which are studied in detail by Sharma et al. (in prep).
    Middle panel: the evolution of the Photon Index during the outburst. 
    Bottom panel: the unabsorbed fluxes in 3-79 keV from \nustar observations in 2014 and 2024, together with the corresponding bolometric X-ray luminosities. The luminosity of the 2024 \nustar observation is estimated based on the joint fitting result, and then a correction for \nustar 3-79 keV flux is calculated and applied to the observations 2014n1 and 2014n2.
    }
    \label{fig:leia_swift_lc}
\end{figure}

Based on the spectral fitting results, the absorption-corrected flux for each \swift-XRT and \LEIA observation was estimated and light curves are shown in Fig.~\ref{fig:leia_swift_lc}.
\LEIA observations of J0520 were significantly affected by the presence of the nearby bright supernova remnant N132D (MCSNR J0525-6938, \citealt{Maggi2016}), resulting in large uncertainties in the soft band. Therefore, we used data in the 1.5-5 keV band for fitting and analysis. The fluxes derived from the \swift-XRT spectra were also calculated within the same energy band.
The fluxes in the 1.5-5 keV range from \nicer observations of J0520 during the same period are also included in Fig.~\ref{fig:leia_swift_lc} for comparison. The comprehensive analysis of the \nicer data is detailed in another study (Sharma et al. in prep).

To estimate the bolometric X-ray luminosity, which is related to the accretion rate and could be used for analysing the spin evolution of the NS, we calculated the 1.5-5 keV flux from the joint fit and then computed the broadband X-ray luminosity $L_{\mathrm{X}}$ by setting the constant for FPMA to unity. The bolometric correction was calculated to be $\sim 6.55$, and applied to estimate the bolometric X-ray luminosity in Fig.~\ref{fig:leia_swift_lc}.

\begin{figure}
    \includegraphics[width=\columnwidth]{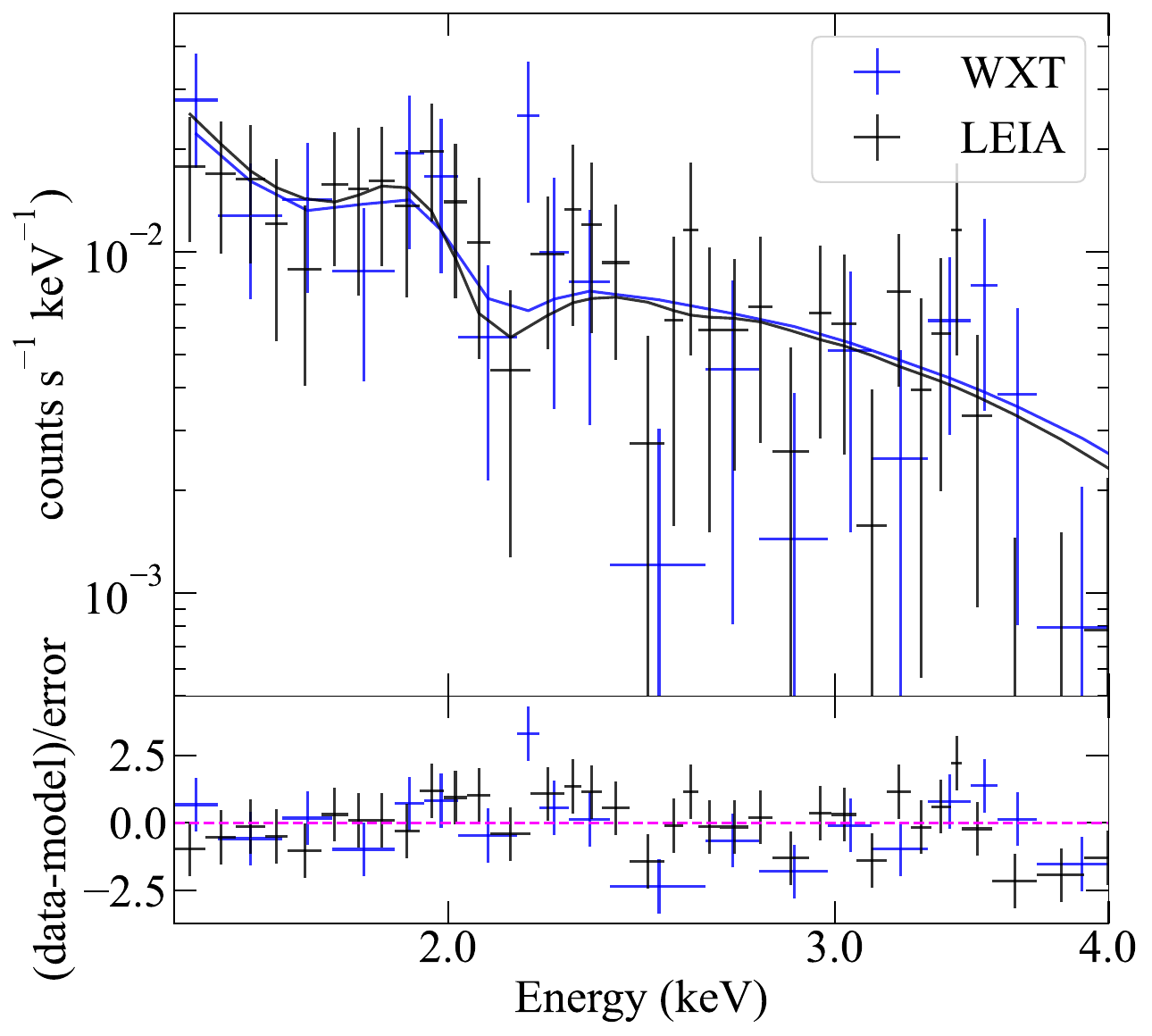}
    \caption{
    The \LEIA and \EP-WXT spectra derived during the same period, along with the best-fit absorbed power-law model. Black and blue cross marks indicate \LEIA and \EP-WXT data, respectively. 
    }
    \label{fig:spec_wxt_leia}
\end{figure}

\begin{figure*}
    \includegraphics[width=2\columnwidth]{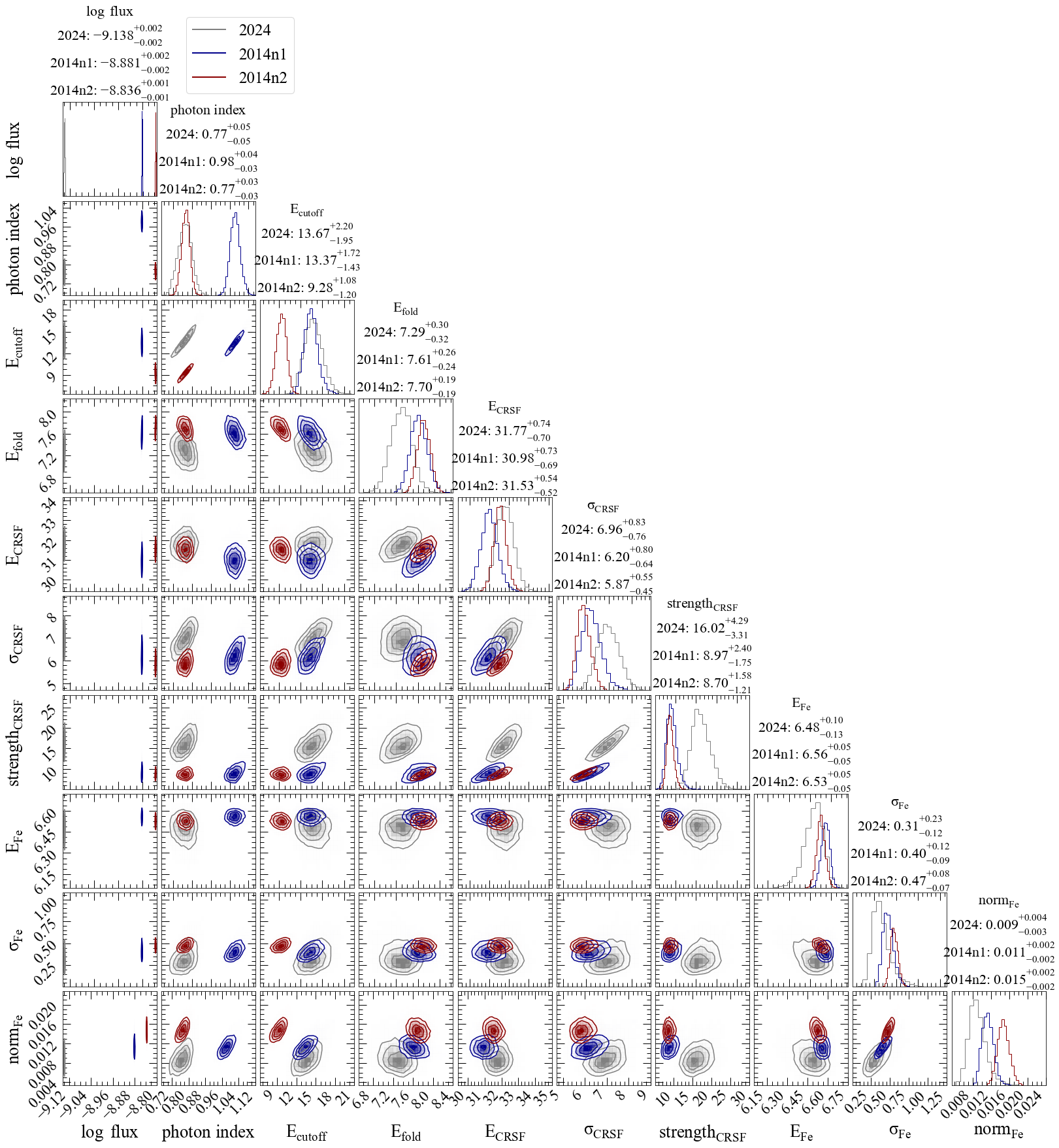}
    \caption{\nustar contours of \texttt{const*tbabs*cflux(powerlaw*fdcut*gabs+gauss)} model for 2024 observation, comparing with observations 2014n1 and 2014n2.}
    \label{fig:contours_nustar_combined}
\end{figure*}

The \nustar data of the 2024 outburst were also compared with the data from the 2014 outburst. The distributions of parameters derived from Markov Chain Monte Carlo (MCMC) simulations are illustrated in Fig.~\ref{fig:contours_nustar_combined}. For the photon index and $E_\mathrm{cutoff}$, which are degenerated with each other, the $E_\mathrm{cutoff}$ from the new observation is near that of 2014n2, while the photon index is closer to 2014n1. The strength of the CRSF also shows variations when compared to the observations in 2014, with the current outburst showing larger line width and depth.

\begin{figure}
    \includegraphics[width=\columnwidth]{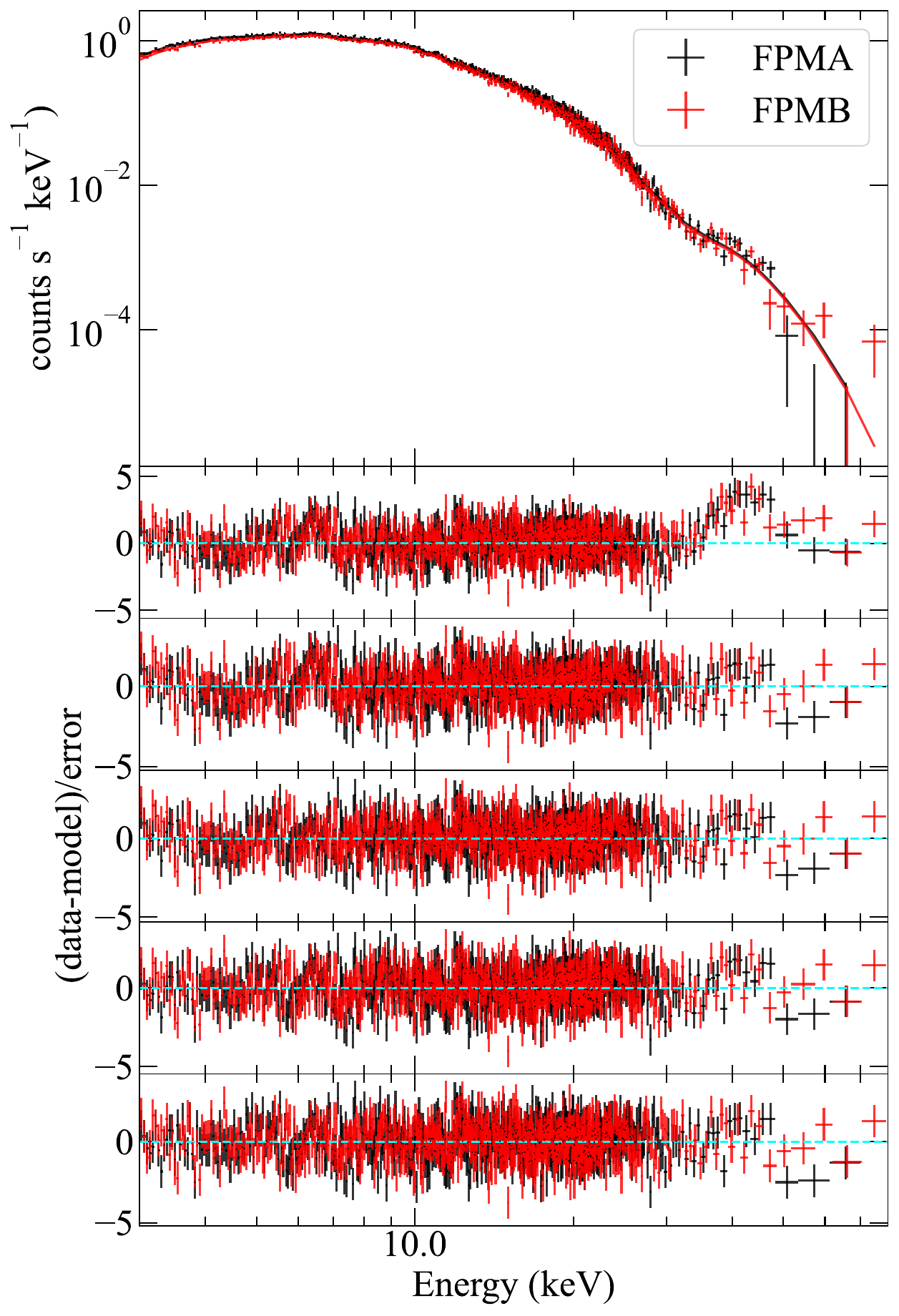}
    \caption{\nustar spectra fit results and residuals for different models. Top panel: spectra with their best-fit \texttt{const*tbabs*cflux(powerlaw*fdcut*gabs+gauss)} models. Black and red cross marks indicate FPMA and FPMB, respectively. Second to sixth panels are residuals corresponding to models in Table~\ref{tab:fit_nustar}. The spectra from the 2014 outburst observed by \nustar appear similar to those from 2024; therefore, only a comparison of the specific parameters is presented in Fig.~\ref{fig:contours_nustar_combined}.}
    \label{fig:fit_nustar}
\end{figure}

\subsection{Timing Analysis}

\subsubsection{Periodic Signal Search} \label{sec:p_search}

\begin{figure}
    \includegraphics[width=\columnwidth]{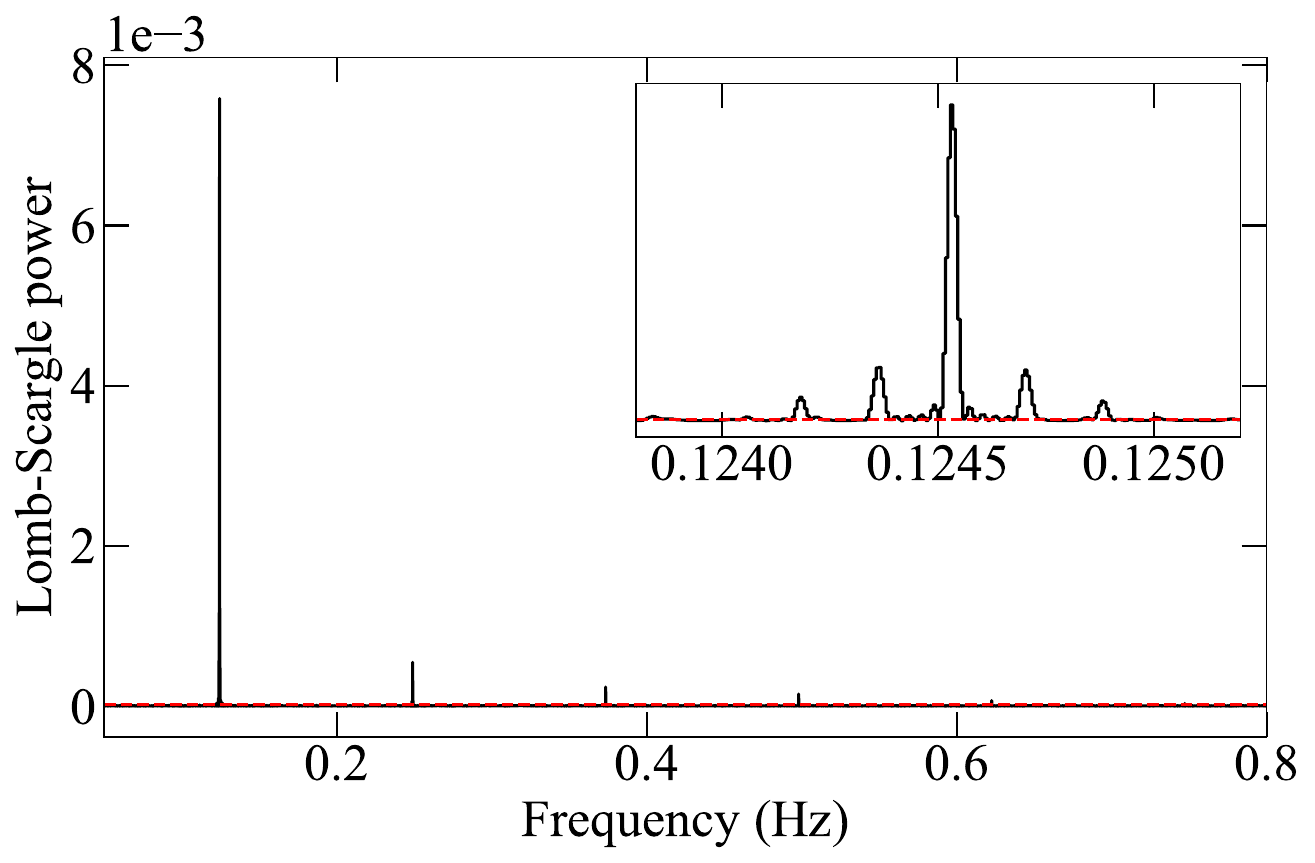}
    \caption{Lomb-Scargle Periodogram of \nustar data. The dashed red line marks the 99.73\% (3$\sigma$) confidence level obtained from the simulation.}
    \label{fig:ls_psd_nustar}
\end{figure}

The \nustar observation with long exposure time ($\sim$19 ks, see Table~\ref{tab:obs_log}) was used to search for any probable periodic signal. The barycenter-corrected event files were used to generate light curves with a time bin of 0.01s for FPMA and FPMB, which were combined for further timing analysis. By using the \efsearch tool, we find a periodic signal at $\sim$8.03 s in the barycenter-corrected data from the \nustar observation, which is close to the value reported by \citet{Vasilopoulos2014} and \citet{Tendulkar2014}.
A significant signal of similar frequency was also identified in the Lomb-Scargle Periodograms of \nustar, as shown in Fig.~\ref{fig:ls_psd_nustar}. 
To estimate the uncertainty of the period, we generated a series of 1000 simulated light curves following the method of \citet{Gotthelf1999} for the \nustar observation. The distribution of measured periods of these simulated light curves corresponds well to a Gaussian function. We derived the $1\sigma$ uncertainty from the standard deviation of this distribution, thus the detected period is $8.029875\pm0.000015$ s.

\begin{figure}
    \includegraphics[width=\columnwidth]{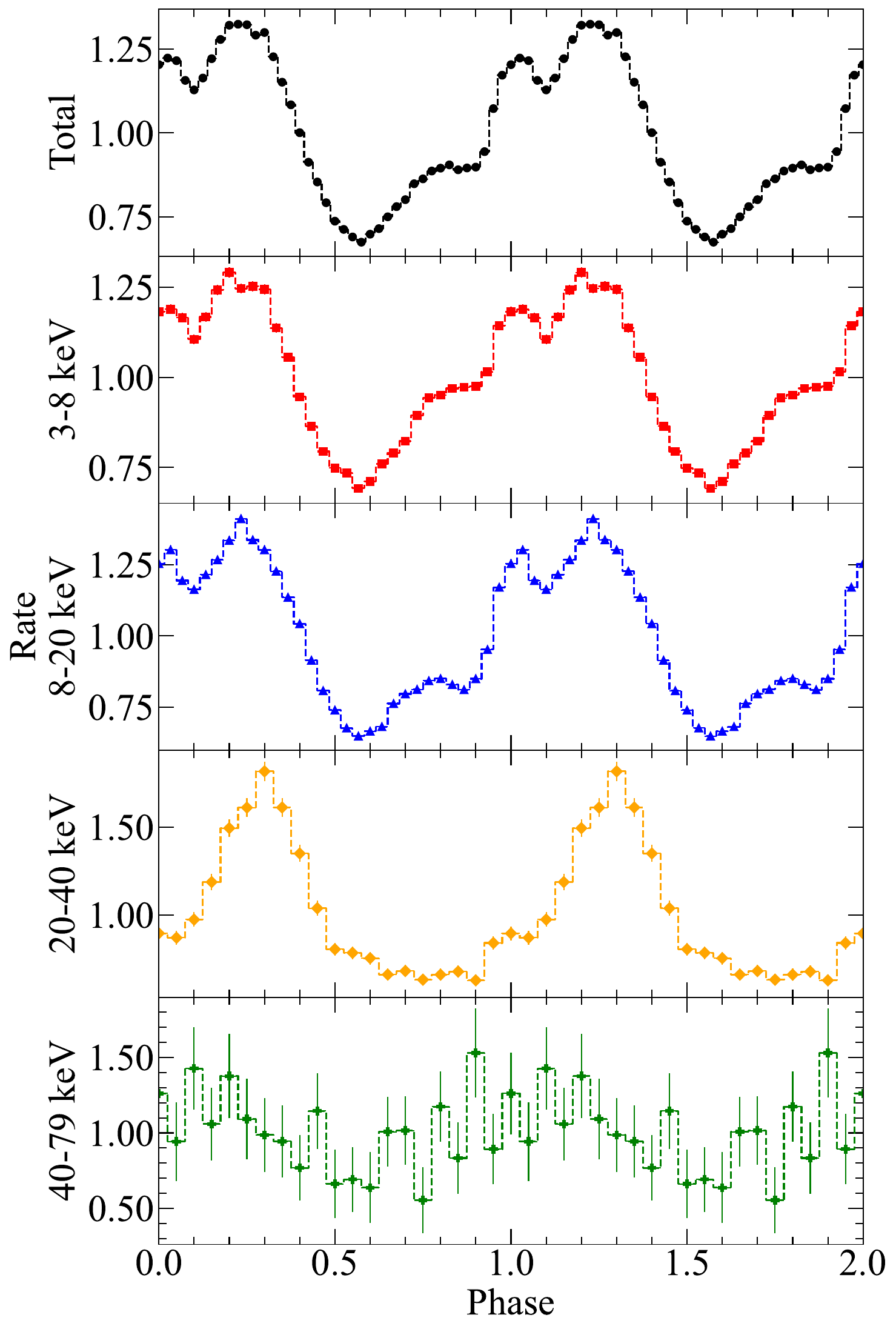}
    \caption{\nustar pulse profile in different bandpasses for 2024 outburst. Top panel: the total pulse profile from 3-79 keV. Second to fifth panels show the 3-8 keV, 8-20 keV, 20-40 keV, and 40-79 keV pulse profiles, respectively. The pulse profiles are plotted with 1-$\sigma$ error bars. 
    }
    \label{fig:profile_nustar}
\end{figure}

The \nustar light curve was folded based on this pulse period with the starting epoch of MJD 60412.0. 
To investigate the pulse features of different energy bands, the counts distribution in the current observation is examined. We first divide the data at 40 keV, where the counts before 40 keV account for 99\% of total observation. The 3-40 keV data was then divided at 20 keV. For the data before 20 keV, we further divided them at 8 keV for the counts in 3-8 keV and 8-20 keV at the same level.
Pulse profiles of 3-8 keV, 8-20 keV, 20-40 keV, and 40-79 keV were generated using the light curves in these different bandpasses, which are shown in Fig.~\ref{fig:profile_nustar} together with the average profile of the total band.

The profiles show dramatic changes in different bands. The total pulse profile is characterized by its multi-peak shape, which originates mainly from the 3-8 keV and 8-20 keV photons. The primary and secondary peaks at phases $\sim$0.3 and $\sim$0.05, which are separated by a single narrow dip, are sharper than the tertiary peak at the phase of $\sim$0.8. In contrast, the 20-40 keV profile only shows a single pulse that rapidly increases and decreases with a maximum at phase $\sim$0.3. It is hard to distinguish pulses in the 40-79 keV profile due to the low count rate above 40 keV.

\begin{figure}
    \includegraphics[width=\columnwidth]{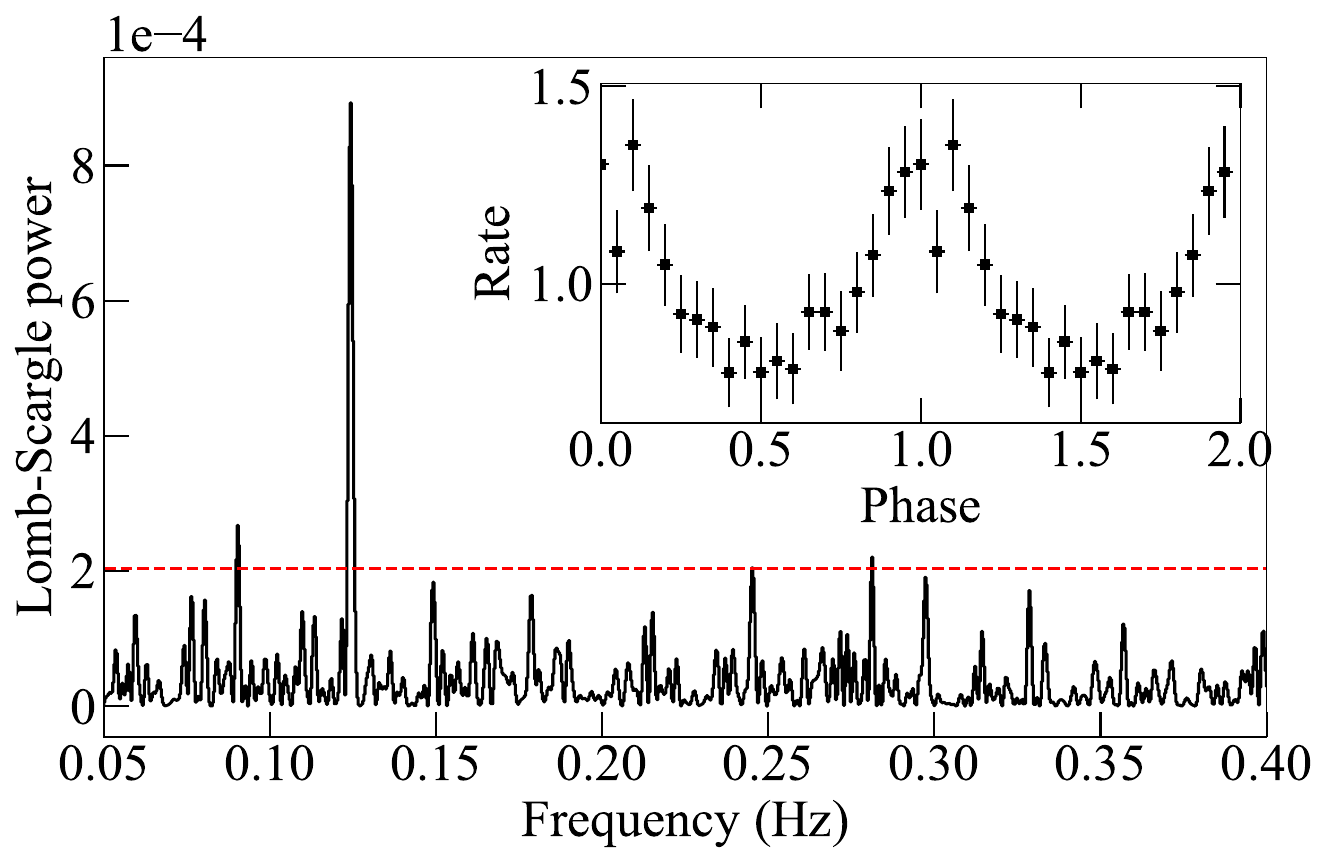}
    \caption{The Lomb-Scargle Periodogram and pulse profile with 1-$\sigma$ error bars of \swift-XRT WT mode second part data from observation s091. The dashed red line marks the 99.73\% (3$\sigma$) confidence level obtained from the simulation.}
    \label{fig:ls_psd_swift_091}
\end{figure}

Given that \swift-XRT WT mode observations also provide data of high time resolution, we also checked the first two \swift observations 00032671090 (briefly referred to as s090 hereafter) and 00032671091 (s091 hereafter) of which the WT mode exposure time is longer than 100s. 
The Lomb-Scargle Periodogram calculated based on the light curve of s090 did not reveal any periodic signals, which may be due to its relatively short exposure time. The observation s091 exposure was divided into 2 parts, the first part lasts for 18 s and the second part lasts for 587 s, with a gap of $\sim$45400s between them. We only use the second one as the main part for further timing analysis. Similar to the \nustar observation, the LS Periodogram of s091 also displays a significant signal at $\sim$0.1245 Hz, as shown in Fig.~\ref{fig:ls_psd_swift_091}.

Considering the duration of 587 s, the periodicity searching on s091 data could be subject to large uncertainty and is difficult to estimate. Consequently, the light curves of s090 and s091 were folded according to the same period measured using data from \nustar, with MJD 20412 and 20416.28 selected as the starting epoch, respectively. Both profiles exhibit at least one pulse, with the results for s090 appearing particularly noisy. Fig.~\ref{fig:ls_psd_swift_091} shows the pulse profile of s091.

\begin{figure*}
    \includegraphics[width=\columnwidth]{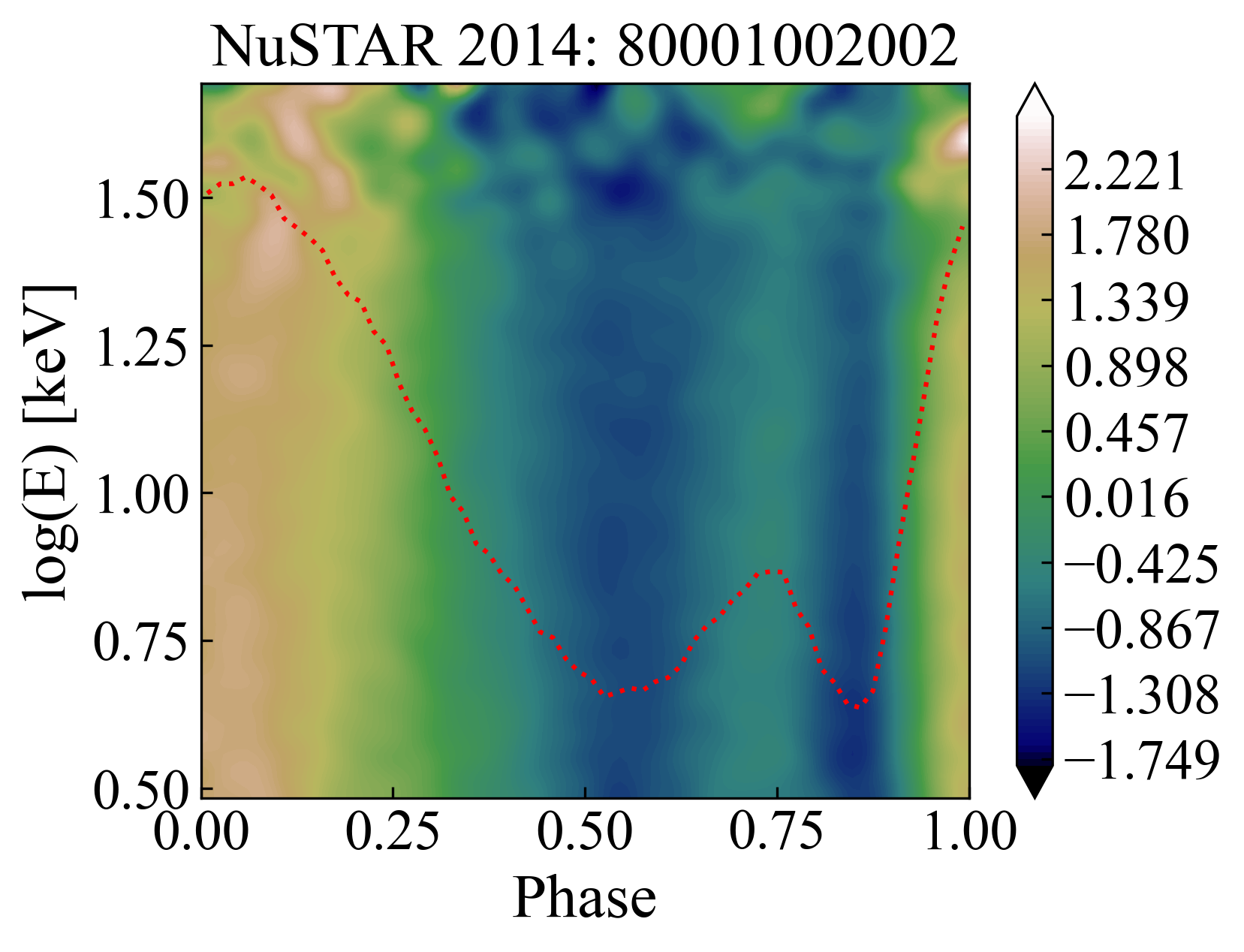}
    \includegraphics[width=\columnwidth]{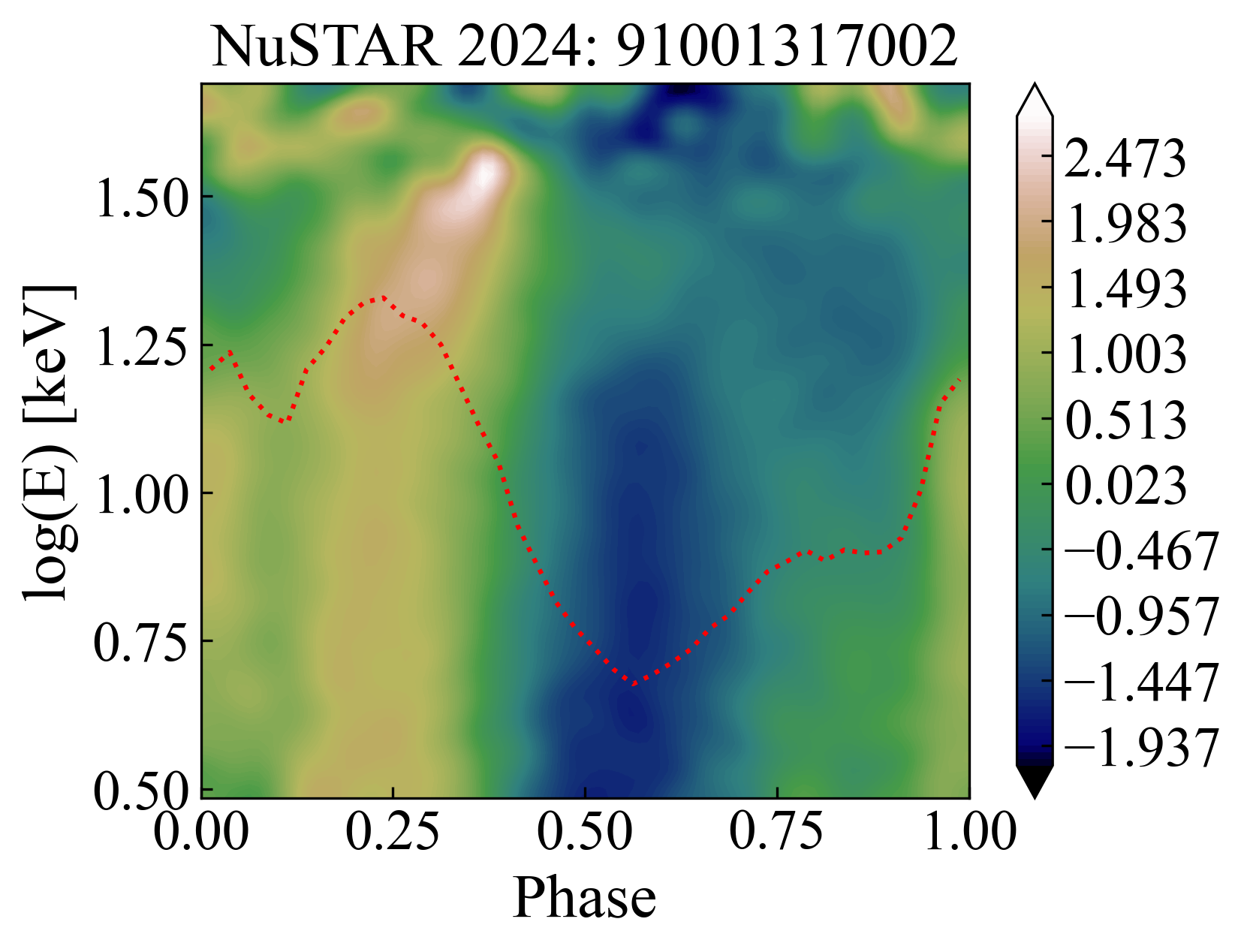}
    \caption{\nustar phase energy heat-maps for 2014 and 2024 observations (i.e. obsids 80001002002 and 91001317002). Each energy bin is normalized by subtracting the average pulse intensity and subtracted by the standard deviation of the energy bin. For clarity, we plot the 3-50 keV pulse profile. We note the sharp dip before the main peak in the 2014 data is not present in the 2024 data, which do however show a more complex peak structure and are more energy dependent. In particular, the 2024 data reveal a change in pulse shape around phase 0.9-1.0 and energy 15-20 keV (see right panel, $\log(E)\sim1.2$). In addition, the main peak drifts from phase $\sim$0.2 at lower energies to phase $\sim$0.35 around 30 keV, while the main trough at phase $\sim$0.6 begins to fill up at higher energies. Nevertheless, both datasets become more noisy above 30 keV where background starts to dominate. 
    } 
    \label{fig:Nu_heat}
\end{figure*}

\subsubsection{Improved Orbital Solution} \label{sec:new_orb}

The 2014 outburst of the system lasted for several binary orbits and enabled detailed modelling of the orbital motion and intrinsic spin-up \citep[see][]{Karaferias2023}. However, the new outburst requires extrapolating the orbital solution over 10 years, yielding a spin-down trend during the 2024 outburst, which is not consistent with a major outburst. This is mainly a result of the limited accuracy of the prior solution, and the data acquired in the new outburst provides an excellent opportunity to improve the measured orbital period of the system.

Keplerian orbits are characterized by five orbital elements: the orbital period ($P_{\rm orb}$), the orbital eccentricity ($e$), the argument of periastron ($\omega$), the projected semimajor axis ($a\sin{i}$ in light-seconds), and the time of mean longitude of 90 degrees ($T_{\rm \pi/2}$) for the orbital phase.
The intrinsic spin-up is associated with the size of the inner radius of the accretion disk and the mass accretion rate which in turn can be tied to the observed luminosity of the system \citep[see examples][]{2016ApJ...822...33P}. 

For the modelling, we followed the Bayesian approach outlined in \citet{Karaferias2023}.
The method employs a nested sampling algorithm for Bayesian parameter estimation, obtaining posterior distributions for both standard accretion torque models and binary orbital parameters. 
To derive the posterior probability distributions and Bayesian evidence, we used the MLFriends nested sampling MC algorithm \citep{2004AIPC..735..395S,2019PASP..131j8005B}, implemented via the {\sc ultranest}\footnote{\url{https://johannesbuchner.github.io/UltraNest/}} package \citep{2021JOSS....6.3001B}. 
This method was applied on the bright outburst of J0520 \citep{Karaferias2023}, while the same method had been used in constraining orbital periods with systems with noisy data \citep[see][]{2022A&A...664A.194V}.

\begin{figure*}
    \includegraphics[width=\columnwidth]{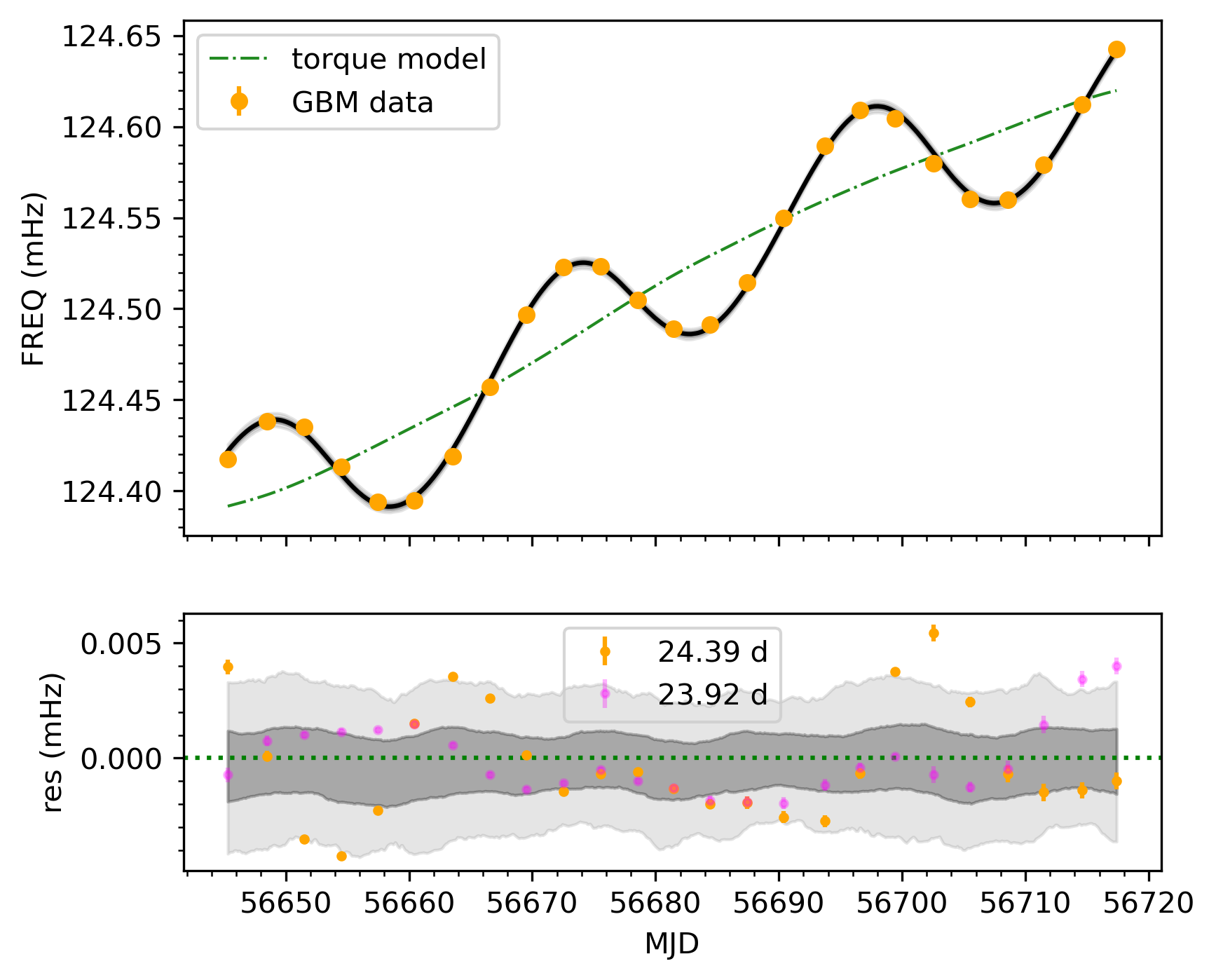}
    \includegraphics[width=\columnwidth]{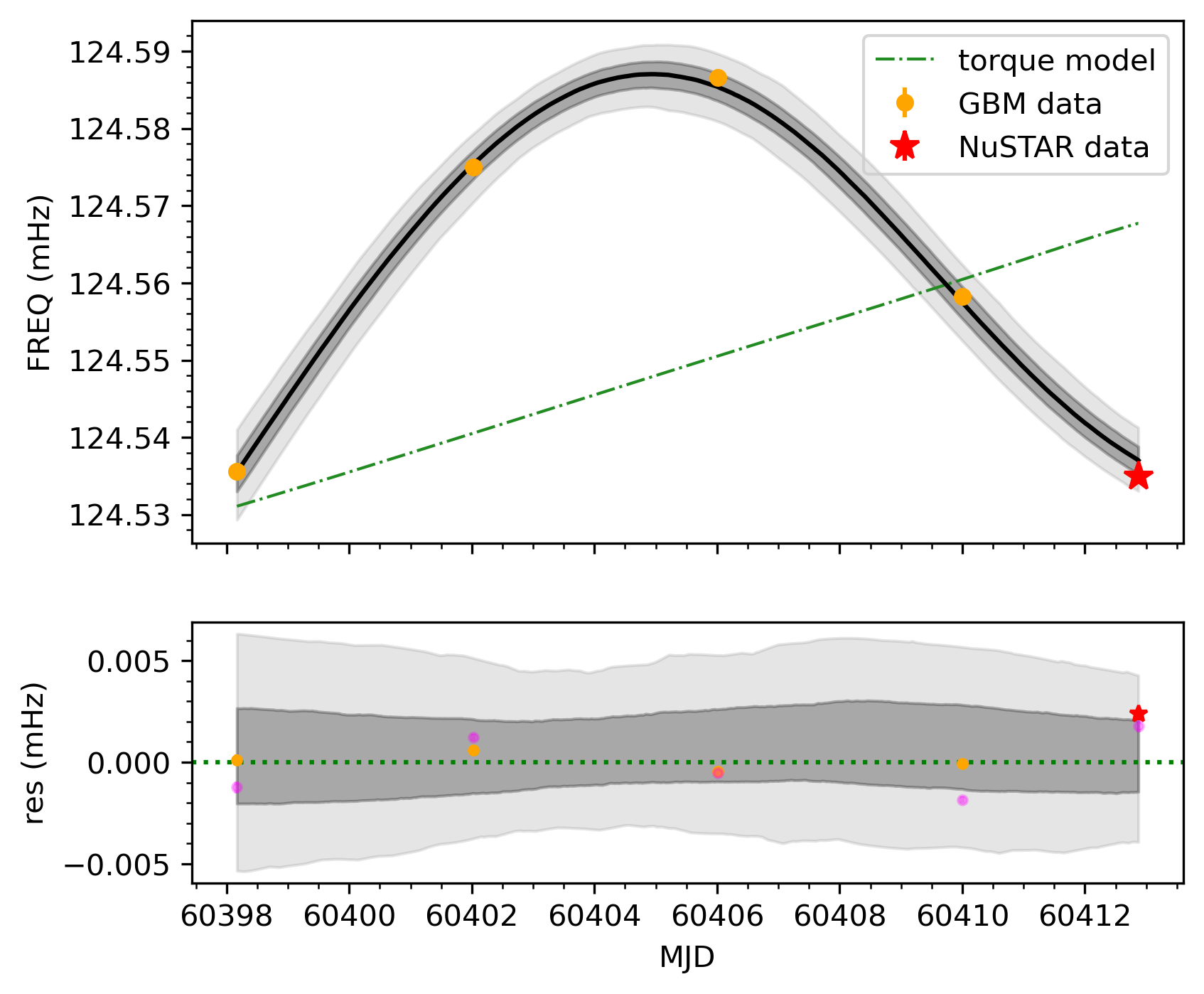}
    \caption{Frequency evolution during the 2014 and 2024 outbursts of J0520 based on the calculation in \ref{sec:new_orb}. 
    Barycentric corrections have been performed for GBM (orange) and \nustar (red) frequency measurements, thus the variability seen is due to binary orbital effects and intrinsic spin-up of the accreting NS. 
    With shaded regions, we plot the prediction bands from the posterior probability distribution using 1 $\sigma$ (dark grey) and 98\% confidence intervals (light grey). In Both panels we also plot the intrinsic spin-up due to accretion based on the adopted torque model (dot-dashed green line). The residuals are estimated based on the deviation from the model derived from the values in Table \ref{RXJ0520orbitals} and for both solutions (orange vs magenta points), however the prediction band is only presented for the 24.3886 d orbital period (i.e. Solution II). }
    \label{fig:time_gbm_period2}
\end{figure*}

For the 2014 outburst, we used the \fermi/GBM frequency measurements and \swift/BAT data as a proxy for the bolometric X-ray luminosity $L_\mathrm{X}$.
For 2024 we used the GBM frequencies (only 4 available data points) and one \nustar measurement, while for $L_\mathrm{X}$ we used data presented in Fig.~\ref{fig:leia_swift_lc}. We assumed that for the accretion material, all dynamical energy is converted to radiation and adopted the canonical NS parameters, i.e. 1.4 M$_{\odot}$ and 12 km.
The magnetic field $B$ is another free parameter of the model\footnote{To avoid any confusion in \citet{Karaferias2023} the magnetic field quoted in tables is the equatorial field, while here we report the polar $B$ field strength, i.e. factor 2 stronger} while for the induced accreting torques we followed \citet{1979ApJ...234..296G}, assuming a ratio of magnetic radius to Alfven radius of 0.5. We also introduced a jump condition for the frequency between the two outbursts, thus we use $F_0$ and $F_1$ as reference frequencies at the start of each outburst. Finally, we introduced an excess noise term $\log{f}$ to account for the systematic scatter and noise of our data not included in the statistical uncertainties of the measurements and model.
Our results for the 2024 epoch are shown in Fig.~\ref{fig:time_gbm_period2}, while the corner plot of the parameters and their values are presented in Fig.~\ref{fig:time_gbm}, and Table~\ref{RXJ0520orbitals}. The evidence $\ln{Z}$ of our fit is also included in the table, 
which gives the marginal likelihood and can be used as a measure of the goodness of fit. The two solutions exhibit only minor differences in the residuals. Only the results of Solution II, in which the orbital period is closer to the reported optical period \citep{{Vasilopoulos2014}}, are presented. A discussion of the two different solutions will be provided in Section \ref{sec:orb_p_discussion}.

\begin{table}
	\centering
	\small
    \setlength\tabcolsep{2pt}
	\caption{Orbital and Torque Model Parameters of J0520.}
	\label{tab:Orbitals}
	\begin{tabular}{lccr} 
		\hline
		Params & Solution I & Solution II& Units\\
		\hline
        $log(B)$ & 11.996 $\pm$ 0.009 & 12.005$\pm$0.014 &G\\
		$e$ & 0.048 $\pm$ 0.012 & 0.05$\pm$0.019 &-\\
		$P_{\rm orb}$ & 23.9188 $\pm$ 0.0007 & 24.3886$\pm$0.0012 &d\\
		$\omega$ & 243 $\pm$ 15 & 256$\pm$27 &$^{o}$\\
		$a \sin i$ & 106.1 $\pm$ 1.3 & 108.3$\pm$2.0 & 1ight-sec\\
		$T_{\rm \pi/2}$ & 56666.47 $\pm$ 0.05 & 56666.17$\pm$0.08 & MJD\\
		$F_{0}$ & 124.3930 $\pm$ 0.0007 & 124.3919$\pm$0.0011 & Hz \\
		$F_{1}$ & 124.5315 $\pm$ 0.0008 & 124.5306 $\pm$0.0013 & Hz \\
		$ln(f)$ & -13.36 $\pm$ 0.15 & -12.81$\pm$0.16 & \\
		$ln(Z)$ & 348.232 $\pm$ 0.548 & 338.549$\pm$0.587 & \\
        \hline
        MJD$_{\rm ref} (F_{0})$ & \multicolumn{3}{c}{56645.3 (fixed)}\\  
        MJD$_{\rm ref} (F_{1})$ & \multicolumn{3}{c}{60398.169 (fixed)} \\
		\hline
	\end{tabular}
	\begin{tablenotes}
      \small
      \item 
    \end{tablenotes}
    \label{RXJ0520orbitals}
\end{table}

\subsubsection{Phase-resolved Measurements}

Based on the detected period and epoch setting in Section \ref{sec:p_search}, we generate several good time intervals (gtis), corresponding to different phase bins, to trace spectral changes during different rotation phases. The photons from the \nustar observation of the new outburst were divided into 10 equal phase bins and then new spectra were extracted and binned to ensure at least 20 photons in each bin, in the same way as the total observation.

The best-fit model of both joint fitting result and \nustar fitting result was used to fit the phase-resolved spectra from FPMA and FPMB in all phase bins. The 20 spectra were fit simultaneously, with FPMA and FPMB spectra in each phase bin tied. The $N_{\mathrm{H}}$ is frozen at the best-fit value of joint fit. Due to the weak intensity of the Fe lines, we fixed the central energy and width of the Gaussian component at the value derived from the phase-averaged spectra, allowing only its normalization to be a free parameter. Preliminary fits show that the $E_{\mathrm{fold}}$ values of the \texttt{fdcut} component do not vary significantly over various phases, and are not related to the other parameters. 
Thus, to better investigate the variations of CRSFs, we also fixed the $E_{\mathrm{fold}}$ parameters at the average value. 
This approach facilitates a more detailed examination of the CRSF behaviour while maintaining constraints grounded in the broader spectral characteristics.

\begin{figure*}
    \includegraphics[width=2\columnwidth]{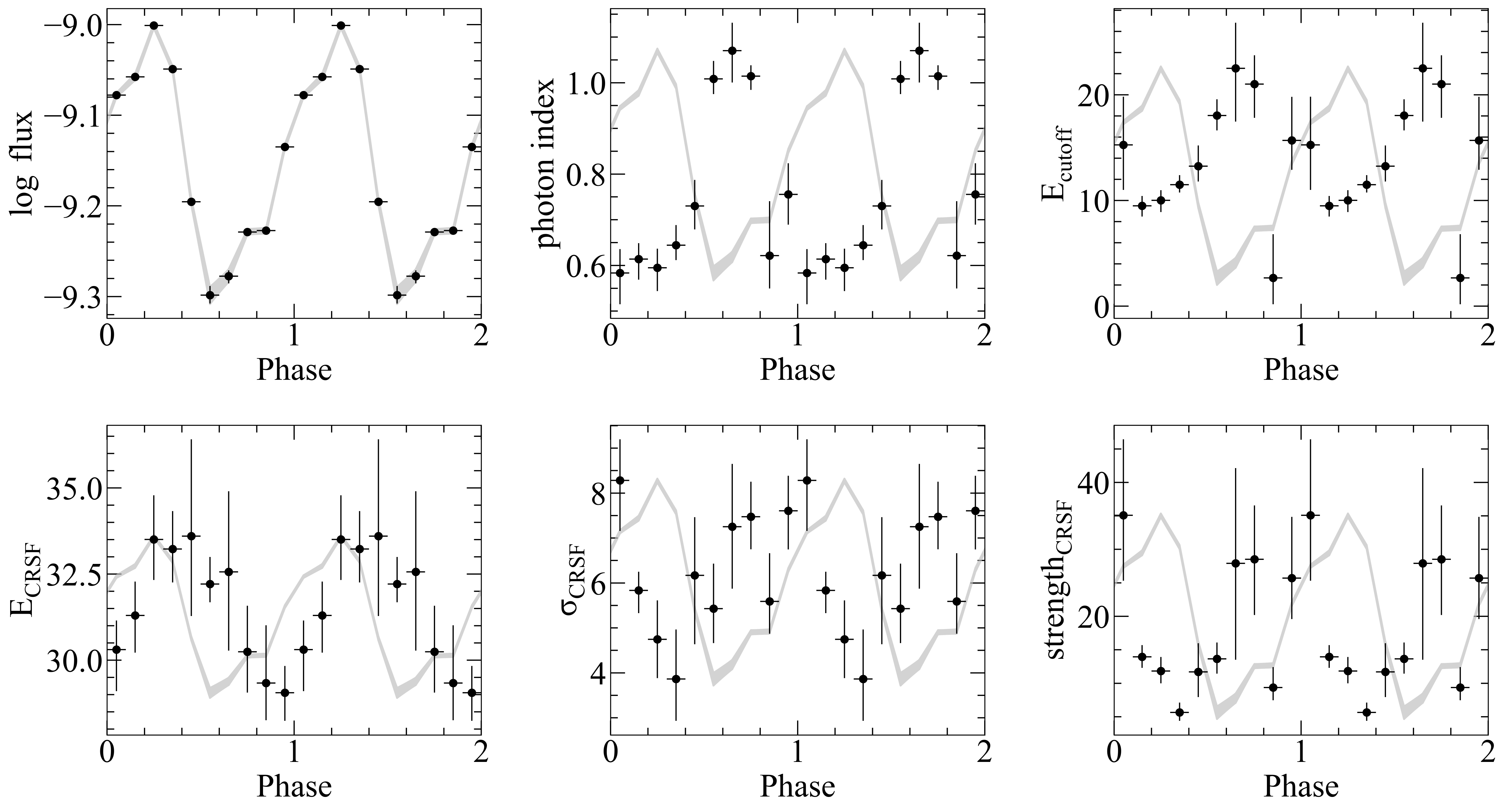}
    \caption{\nustar model parameters as a function of phases. Errors represent 90\% confidence intervals from MCMC (200,000 step MCMC run with a 200,000 step burn-in). The 90\% error region shown as a grey shadowed area comes from random sampling under distributions of $\log$ flux and is re-scaled in other panels to indicate the general flux variation. 
    }
    \label{fig:par_phase_nustar_group_1}
\end{figure*}

In order to estimate the uncertainty of each parameter and measure the variation in different phase bins, we run a MCMC simulation of $2\times 10^5$ steps with a $2\times 10^5$ step burn-in. The results are shown in Fig.~\ref{fig:par_phase_nustar_group_1}.

\section{Discussion}\label{sec4}

\subsection{Orbital and optical periodicities}\label{sec:orb_p_discussion}

\begin{figure*}
    \includegraphics[width=0.32\textwidth]{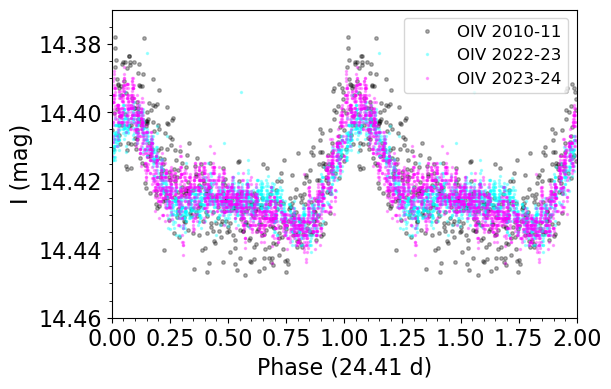}
    \includegraphics[width=0.32\textwidth]{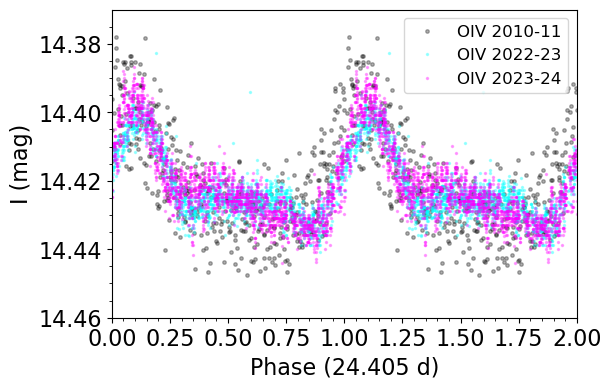}
    \includegraphics[width=0.32\textwidth]{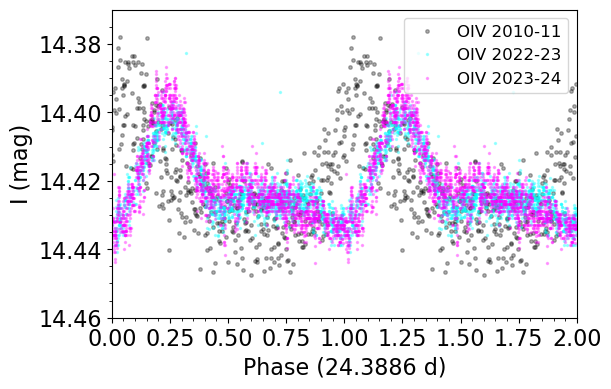}
    \caption{Optical profile folded for periods derived by the timing analysis. Data obtained prior to 2020 are marked with black points while 2 more recent OGLE IV epochs are marked with colors. In the left panel we plot all available OGLE data, while in the other two we only plot data before 2011 and after 2022 so phase drifts are easier to notice. All data are de-trended and 2 epochs of major outbursts have been removed for clarity.}
    \label{fig:OGLE_Profile}
\end{figure*}

In this work we have derived an updated orbital solution based on the spin period evolution during two major outbursts separated by 10 years. 
Prior to the 2024 outburst, the best estimates of the orbital period were 23.93$\pm0.07$ d \citep{2014ATel.5856....1K} and 23.97$\pm$0.06 d \citep{Karaferias2023}. Our solution derived similar orbital parameters within uncertainties apart for the orbital period. We note that there is a degeneracy in the solutions presented in Section \ref{sec:new_orb} concerning the orbital period, that has to do with the number of orbital cycles between the two outbursts. In terms of goodness of fit (see log Evidence) a period of 23.91 d might be preferred, however, a period of 24.38 d is closer to the reported optical period \citep[i.e. 24.43 d][]{Vasilopoulos2014}. 

We analysed the publicly available OGLE data that cover 14.2 years\footnote{OGLE XROM project: \url{https://ogle.astrouw.edu.pl/ogle4/xrom/rx-j0520.5-6932.html}} to provide an updated comparison with the optical period. We de-trended the OGLE I band data and evaluated a Lomb-Scargle periodogram for the complete data set as well as subsets of data \citep[see application][]{2021MNRAS.503.6187T}. The optical period varied between 24.4-24.45 d, while by folding the data the optimal alignment of all epochs was obtained for a period of 24.41 d, as shown in Fig.~\ref{fig:OGLE_Profile}. We note that for the period derived from the pulse evolution there is a gradual offset (0.15 in phase) between optical data obtained over the 14-year period.
We also attempted to fit the pulse period evolution with a prior for the orbital period between 24.4-24.45 days and the solution could not fully align the 2014 and 2024 data.
More complicated models like adopting a variable orbital period are not explored. Nevertheless, the shift in the optical data would require a change of 0.05 d over 14 years or, about $10^{-5}$ d/d or 5 minutes per year which is quite high to explain in terms of binary orbital evolution even at extreme accretion rates \citep{2021ApJ...909...50V,2021arXiv211203779K}. Thus the optical period being a beat period of the orbital period would be a more realistic scenario. 

The orbital period derived from outburst modulations could be influenced by the precession of the Be disk. Orbital phase shifts or even jumps could also happen (e.g., \citealt{Wilson2002}). However, the optical and X-ray data of J0520 during the two outbursts are well-aligned, suggesting a possible connection between them.
If the two periods ($\sim$24.39 d vs $\sim$24.41 d) are connected via the same beat effect then it would suggest a super-orbital period of about 76 years. On the other hand if Solution I is adopted, then the beat period of the optical ($\sim$24.41 d) and orbital period ($\sim$23.92 d) should be $\sim$3.2 years, indicating that the 2014 and 2024 outbursts are separated by 3 such cycles. At this point, it is not possible to strongly argue in favour of either orbital solution, as a third major outburst would enable improvement of the orbital solution. 
Our conclusion is that the periods search from optical data and X-ray data (pulse timing) cannot agree on a single solution, but there can be a mismatch of as low as 0.02 days between the two periods. 

\subsection{Optical and X-ray Alignment}
\citet{Edge2004} revealed that J0520 experienced an outburst detected in both optical and X-ray in 1995. The data from OGLE, \swift-XRT, \LEIA and \EP also provide monitoring of the system's outbursts in both optical and X-ray bands covering the years 2014 and 2024. 
The improved period of 24.39 days\footnote{Adopting either the 24.39 d or 23.92 d orbital period has no effect in our results or in Fig.~\ref{fig:multiband}.}, derived from the X-ray data, matches perfectly with the OGLE light curve not only in the last 2 years but also for the 10 years, indicating a relatively stable periodicity of the system. 
The sudden changes in optical flux correspond well with the X-ray outbursts in both outbursts, while the rise and peak of the X-ray flux precede those of the optical flux. 
The major outburst of BeXRB is associated with changes in the circumstellar disk of the Be star, which can trigger a high mass accretion rate onto the NS, leaving traces in both optical and X-ray bands (e.g. \citealt{Chhotaray2023}). Additionally, the simultaneous rises in optical and X-ray, similar to the 2017-2018 giant outburst of Swift J0243.6+6124, points to an alternative explanation of X-ray irradiation of the Be-star disk (\citealt{Alfonso-Garzon2024}, Vasilopoulos in prep).

\subsection{Spin Evolution}

Many BeXRBs show spin-up during major outbursts (e.g. \citealt{Bildsten1997, Wilson2008}), which is associated with the accretion-powered nature of the NS. The spin evolution of a BeXRB during a major outburst is considered to be related to various patterns of torque applied to the NS. On the one hand, the system with a luminosity exceeding the Eddington limit accretes material efficiently, resulting in a torque that leads the NS to spin-up. On the other hand, the coupling of accretion flows with magnetic field lines tends to spin-down the NS. Consequently, the total torque coming from both sides changes the spin period, which has been described in many torque models (e.g. \citealt{1979ApJ...234..296G, 1995ApJ...449L.153W}). For extragalactic systems located in the LMC and SMC we can only perform detailed torque modelling during major outbursts \citep[see applications][]{2017MNRAS.471.3878T,2020MNRAS.494.5350V}, whereas for these outbursts we typically observe the systems away from equilibrium.

The torque model proposed by \citet{1979ApJ...234..296G} can be applied to magnetic neutron stars accreting matter from a disk, especially in a strong accreting regime (as reviewed and highlighted by \citealt{Bozzo2009}). For J0520, its spectral characteristics during the outburst are consistent with those of a strongly accreting source.
By using data from both the 2014 and 2024 outbursts with the torque model from \citet{1979ApJ...234..296G}, the orbital parameters are better constrained. Applying this to new outbursts allows for an estimation of intrinsic spin-up during the outburst period.
Our findings indicate that the behaviour during the outburst is consistent with an accreting regime away from equilibrium and a slow rotator, where the magnetosphere is much smaller than the co-rotation radius.
For the particular torque model, we find a magnetic field B around 10$^{12}$~G, which is consistent with the findings of \citet{Karaferias2023}. Another interesting finding is that between the major outbursts, the spin period of the system remained almost constant, increasing from $\sim$8.026 to $\sim$8.03 over a period of 10.3 years, yielding a spin-down of 4.4$\times10^{-4}$ s/yr (see Fig.~\ref{fig:time_gbm_period2}).
This suggests that the interaction between matter and the electromagnetic field is stronger than the accretion of matter. By neglecting the spin-up torque from the accretion flow during the spin-down process, a lower limit of the magnetic field strength of B $\ge 7\times 10^{11}$~G can estimated, which is consistent with the strength measured by the CRSF \citep[see][]{Wang2021}.

\subsection{Cyclotron Resonant Scattering Feature}

Cyclotron resonance scattering features (CRSFs), which have been observed in many HMXB systems, are critical phenomena that offer a direct window into the magnetic environments of NS. 
The physics underlying CRSFs is related to the quantized energy levels of electrons in the presence of strong magnetic fields of NS, known as the Landau levels. The scattering cross-section of electrons resonates at the discrete energies and gets enhanced to high values. Meanwhile, due to the thermal broadening effects of Landau levels, the photons will be absorbed and contribute to the transition of electrons when they are of an energy close to the difference between two electron Landau energy levels.
The energy at which these absorption lines occur directly correlates with the magnetic field strength \citep{Trumper1978, Nagel1981}, thereby enabling precise measurements of NS magnetic fields (for reviews, see e.g. \citealt{Staubert2003, Heindl2004, Caballero2012, Maitra2017, Staubert2019}).

Both the joint spectral fitting and \nustar spectra show the presence of a significant CRSF with a centroid energy of $\sim$32 keV, which is similar to that observed during the outburst in 2014. CRSFs detected both during the 2014 and the current outburst are statistically significant, with the simftest indicating a 100\% probability of significance. The magnetic field strength can be estimated to be $\sim 3.6\times 10^{12}$ G using the 12-$B$-12 rule \citep{Schonherr2007}, several times the value obtained according to the torque model.
The critical luminosity for a typical NS $L_{\mathrm{crit}}$, which corresponds to the magnetic field, could then be estimated to be $5.9 \times 10^{37}$ erg s$^{-1}$ \citep{Becker2012}. However, the CRSF of the new outburst shows a higher line width ($\sigma$) and depth (strength), which result in an optical depth of $0.9_{-0.3}^{+0.4}$, close to the value of observation 2014n2.

The X-ray luminosity of J0520 during major outbursts exceeds $L_{\mathrm{crit}}$, implying that it is dominated by radiation pressure near the stellar surface. The energy of the CRSF at this super-critical regime is considered to be negatively correlated with luminosity \citep{Becker2012}. However, when compared on a larger timescale, the anti-correlated behaviour does not appear in the case of J0520. Compared to 2014n2, the luminosity of the current outburst dropped by ~50\%, but the CRSF energy almost stays the same value within measured uncertainties.
Considering the correlation between various spectral parameters (see Fig.~\ref{fig:contours_nustar_combined}), the width and depth of the CRSF component may influence the estimates of its centroid energy. Additionally, the CRSF energy can also be highly dependent on the precision of the continuum modelling.
Nevertheless, the comparison of expected behaviour and measured properties between different outbursts is challenging. For example in the case of SMC X-2 the CRSF energy measured in two different major outbursts at exactly the same luminosity level was different by 2 keV \citep[keV 29.5 vs 31.5 keV, see][]{2023MNRAS.521.3951J}.
Over a timescale of more than a decade, some sources show long-term evolution in the CRSF energy \citep[e.g.][]{Staubert2014, La_Parol2016, Ji2019}.
For other sources like 1A 0535+262 \citep{Kong2021, 2024MNRAS.528.7320S} and V0332+53 \citep{2016MNRAS.460L..99C} there appears to be an energy drift of CRSF energy with time within the same outburst. Interestingly, the opposite effect was been observed in these two systems, with the line energy increasing with time for 1A 0535+262 and dropping for V0332+53. 
These effects could indicate a complex and evolving way of coupling between magnetic field lines and the disk or a magnetic field burial by an advection mechanism. The study of these effects could be the point of emphasis in future X-ray missions like HEX-P \citep{2023FrASS..1092500L}.

\subsection{Pulse Profile}
The pulse profile from \nustar data shows a multi-peak shape and energy-dependent pattern, which have been discovered in many X-ray pulsars (e.g. \citealt{dalFiume1988, Ray2002, Kreykenbohm2008}). Two out of three peaks in the profile disappear when the energy is larger than 20\,keV. As shown in Fig.~\ref{fig:Nu_heat}, the observed profile during the current outburst has a different shape from that of the previous outburst in 2014, and exhibits more significant energy dependence, indicating a varying behaviour instead of a stable pattern (see Fig.~\ref{fig:Nu_heat}). A similar variable pulse shape was also observed from SAX J2103.5+4545 \citep{CameroArranz2007}. Despite the pulse profile is suggested to have no clear correlations with system parameters such as luminosity, magnetic field strength, spin period, and orbital period \citep{Alonso-Hernandez2022}, it provides valuable insights into the physical processes in the accretion column. The narrow dip occurring in the pulse profile in 2024 may be attributed to the interception of our line of sight by the accretion stream from the inner accretion disk to the magnetic poles \citep{Cemeljic1998, Paul2017}.

\begin{figure}
    \includegraphics[width=\columnwidth]{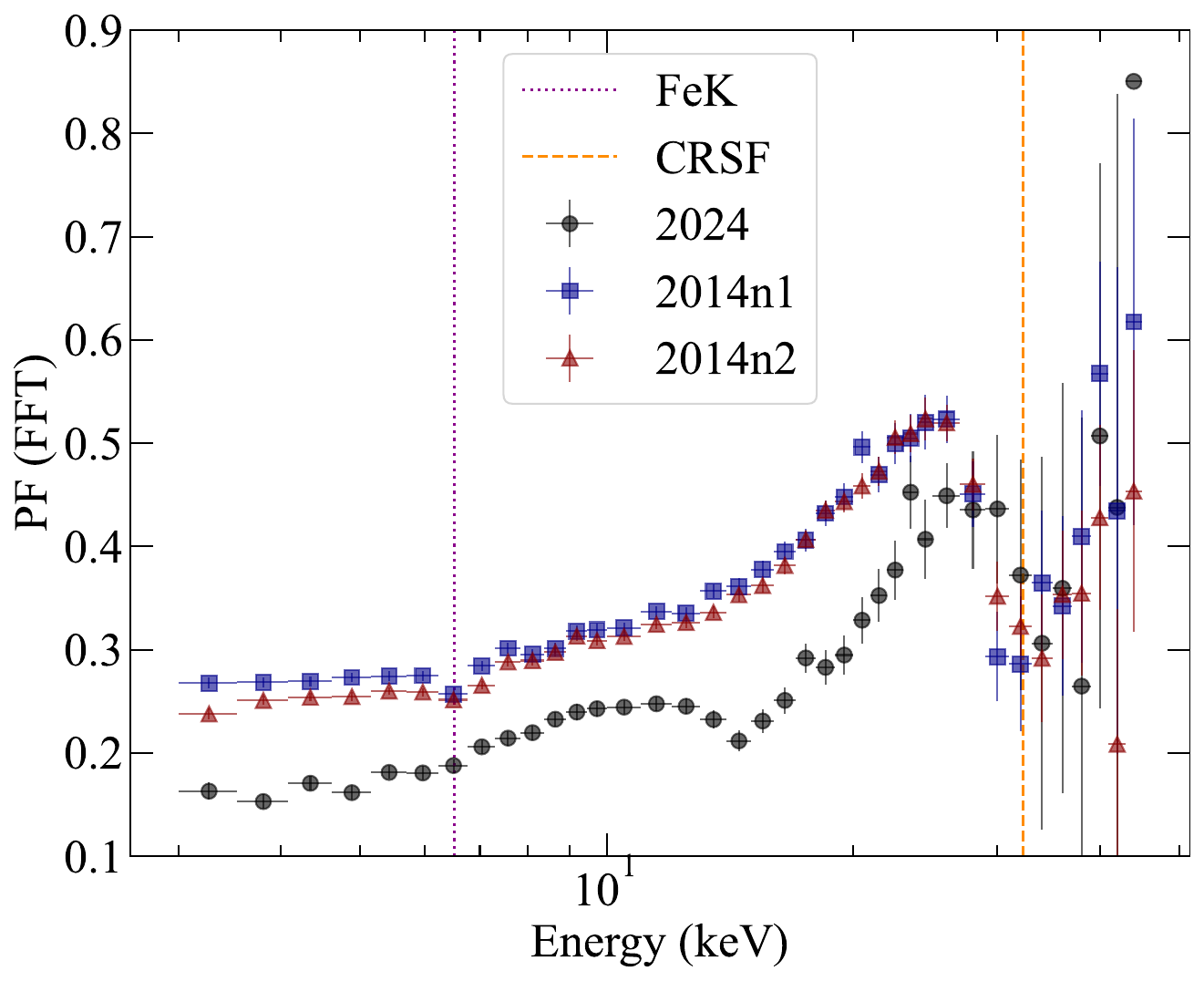}
    \caption{\nustar pulsed fraction as a function of energy for the 2014 and 2024 \nustar observations.
    }
    \label{fig:pf}
\end{figure}

We also calculated the pulsed fraction (PF), which refers to the magnitude of the pulsed component relative to the total emission. It has been known that PF is correlated with energy, and the shape changes in connection to some characteristic features such as the CRSF and the Fe line (e.g. \citealt{Ferrigno2009, Lutovinov2009, Tsygankov2010, Wang2022, Tobrej2023}). The PF has already been recognized as a tool for detecting the presence of certain features \citep{Ferrigno2023}. 
To investigate the energy dependence of the PF that was evident in Fig.~\ref{fig:Nu_heat} we followed \citet{Ferrigno2023} and computed the PF by adopting the fast Fourier transform (FFT) methodology described in their paper. We computed PF$_{\rm FFT}$ for 2014 and 2024 observations and the results are plotted in Fig.~\ref{fig:pf}. In both observations the PF increases with energy up to $\sim$25\,keV where there is a drop centered around the CRSF. However, due to background noise, we cannot resolve the PF evolution above 30 keV. In both the 2014 data we find a drop in PF around the Fe $k\alpha$ line (6.4 keV), which is more evident in the first observation. However, this feature is not evident in the 2024 data, perhaps due to lower statistics, or decreased pulsed intensity.
The decrease of PF around the Fe $k\alpha$ line is consistent with reprocessing away from the NS surface \citep{2004ApJ...614..881H}. 
The intriguing finding is that in the 2024 data, we notice a decrease of PF around 15 keV that matches the energy where the pulse profile changes (see Fig.~\ref{fig:Nu_heat}). The behaviour of the 2024 data is quite similar to 4U 1626-67 that was observed with \nustar \citep[see obsid 30101029002 in][]{Ferrigno2023}. To our knowledge, this is the first time this decrease in PF is observed in an extragalactic source.

In the phase-resolved spectral analysis, the visibility of the secondary and tertiary peaks is suppressed due to the selection of phase bin size and boundaries, although the overall trend is consistent with the folded light curve in Fig.~\ref{fig:profile_nustar}. Except for flux variations, Fig.~\ref{fig:par_phase_nustar_group_1} also shows variations in the continuum and the CRSF component as a function of the rotation phase. Specifically, the photon index and $E_{\mathrm{cutoff}}$, as two parameters with significant correlations (see Fig.~\ref{fig:contours_nustar_combined}), indicate an inverse relation with flux. When the flux increases with phase, they decrease, and vice versa. 
In other words, J0520 appears softer between the peaks of the pulse profile, a characteristic commonly observed in similar systems (e.g. XTE J1946+274, see \citealt{Maitra2013}; 4U 1907+09, see \citealt{Varun2019}), likely indicating phase-dependent modulation in the emission properties.
On the other hand, the $E_{\mathrm{CRSF}}$ profile shows a lag of $\Delta \phi \approx 0.3$ compared to the flux, which is similar to the small lags observed in other objects (e.g. Cen X-3, see \citealt{Suchy2008, Yang2023}; GX 301-2, see \citealt{Suchy2012}). 
There may be an inverse relationship between the width and depth of the CRSF and $E_{\mathrm{CRSF}}$, although no conclusions can be drawn due to the high level of uncertainty.

\section{Conclusions} \label{sec5}


Based on observations from several instruments including \LEIA, \EP, \swift, \nustar and \fermi, our study provides spectral and timing analyses of the recent 2024 outburst of J0520 and investigates its behaviour in optical and X-ray within a long time span of a decade.

\EP detected the earliest brightening on March 22. The high-cadence sampling by \LEIA during the flare suggests that the peak of the outburst occurred around the 29th, allowing observation of the flux evolution during the event.

Joint spectral fitting using \swift and \nustar data, as well as individual fits of observations, indicate that despite the luminosity decreasing by 50\% compared to the previous outburst, the spectral shape remains generally unchanged. The observed CRSF energy is almost the same as the previous outburst in 2014. Notably, the Fe line intensity has significantly weakened. 

By modelling the accretion torque and orbital parameters of the system with the X-ray data from 2014 and 2024, we have improved the orbital period to 24.39 d, allowing us to derive the intrinsic spin-up trend during the outbursts. However, the orbital period obtained from X-ray data is not completely consistent with that from optical data, which may need a new major outburst to further improve the solution.

The \nustar data from the two outbursts, separated by a decade, revealed variations in the pulse profile, that in 2024 a more complicated structure and energy dependence are observed. PF of the current outburst, although showing no clue of drop near the Fe line which is evident in PF of 2014, indicates a new significant decrease around 15 keV. There are also drops in PF around the CRSF in both 2014 and 2024 data. J0520 can then be reported as the first extragalactic source that shows such decreases in PF.

Additionally, various spectral parameters show different correlations with the spin phase. The change of flux with the rotation phase is aligned with the opposite tendency of photon index and cutoff energy, while the energy of CRSF suggests a lag compared to the flux.

\section*{Acknowledgements}

This work is based on data obtained by Einstein Probe, a space mission supported by Strategic Priority Program on Space Science of Chinese Academy of Sciences, in collaboration with ESA, MPE and CNES (Grant No. XDA15310000), the Strategic Priority Research Program of the Chinese Academy of Sciences (Grant No. XDB0550200), and the National Key R\&D Program of China (2022YFF0711500). 
This work is also based on the data obtained with LEIA, a pathfinder of the Einstein Probe mission, which is supported by the Strategic Priority Program on Space Science of Chinese Academy of Sciences (Grant Nos. XDA15310000, XDA15052100)
This research has made use of data from the NuSTAR mission, a project led by the California Institute of Technology, managed by the Jet Propulsion Laboratory, and funded by the National Aeronautics and Space Administration. Data analysis was performed using the NuSTAR Data Analysis Software (NuSTARDAS), jointly developed by the ASI Science Data Center (SSDC, Italy) and the California Institute of Technology (USA).
We thank \nustar PI for approving our DDT observation and the \nustar Science Operations Center for scheduling and carrying out the observation. We also thank the \swift team for approving and performing our Target of Opportunity observations.

HNY acknowledges support from China Scholarship Council (No.202310740002). 
GV acknowledges support from the 1432 Hellenic Foundation for Research and Innovation (H.F.R.I.) through the project ASTRAPE (Project ID 7802). 
A.B. acknowledges SERB (SB/SRS/2022-23/124/PS) for financial support and is also grateful to the Royal Society, United Kingdom.
We acknowledge the support by the National Natural Science Foundation of China (Grant Nos. 12103061, 12203071).
For the pulsed fraction analysis the authors adapted the code developed by \citet{Ferrigno2023} available from \url{https://gitlab.astro.unige.ch/ferrigno/nustar-pipeline}.

\section*{Data Availability}
We made use of publicly available OGLE data from the X-ray monitoring project \url{https://ogle.astrouw.edu.pl/ogle4/xrom/xrom.html}. X-ray data are available through the High Energy Astrophysics Science Archive Research Center: \url{heasarc.gsfc.nasa.gov}. The data presented in the tables and figures of the paper are available upon reasonable request.
 



\bibliographystyle{mnras}
\bibliography{rxj0520} 



\appendix


\section{Information about the observations and spectral fits}

\begin{table*}
\caption{Observational log.}
\label{tab:obs_log}
\begin{threeparttable}
\begin{tabular}{lllllrrr}
\hline
ObsID       & Instrument     & Start Time (UTC)    & Stop Time (UTC)     & MJD   & Exp. (s) & Rate (cts/s) & HR $^{\mathrm{a}}$ \\ \hline
00032671090 & \swift-XRT (PC)& 2024-04-13 00:30:39 & 2024-04-13 00:37:53 & 60413 & 434      & 2.9          & 1.3    \\
            & \swift-XRT (WT)& 2024-04-13 00:28:52 & 2024-04-13 00:30:38 & 60413 & 105      & 5.7          & 1.4    \\
00032671091 & \swift-XRT (PC)& 2024-04-16 01:09:23 & 2024-04-16 01:20:52 & 60416 & 690      & 3.2          & 1.5    \\
            & \swift-XRT (WT)& 2024-04-16 01:09:11 & 2024-04-16 13:55:56 & 60416 & 605      & 6.5          & 
1.4    \\
00032671093 & \swift-XRT (PC)& 2024-04-30 04:53:18 & 2024-04-30 09:37:54 & 60430 & 537      & 2.4          & 1.2    \\
00032671094 & \swift-XRT (PC)& 2024-05-01 23:10:00 & 2024-05-01 23:27:54 & 60431 & 1073     & 2.8          & 1.8    \\
00032671095 & \swift-XRT (PC)& 2024-05-07 00:59:13 & 2024-05-07 21:44:53 & 60437 & 1750     & 2.8          & 1.6    \\
00032671096 & \swift-XRT (PC)& 2024-05-14 05:02:04 & 2024-05-14 08:18:53 & 60444 & 1204     & 2.8          & 1.6    \\
00032671098 & \swift-XRT (PC)& 2024-05-17 05:30:02 & 2024-05-17 05:39:53 & 60447 & 592      & 3.0          & 1.8    \\
00032671101 & \swift-XRT (PC)& 2024-06-04 00:07:13 & 2024-06-04 11:16:53 & 60465 & 1259     & 2.5          & 1.4    \\
00032671102 & \swift-XRT (PC)& 2024-06-07 05:35:21 & 2024-06-07 05:48:54 & 60468 & 812      & 2.5          & 1.4    \\
00032671103 & \swift-XRT (PC)& 2024-06-11 01:13:05 & 2024-06-11 21:43:53 & 60472 & 1698     & 2.4          & 1.4    \\
00032671104 & \swift-XRT (PC)& 2024-06-14 02:58:04 & 2024-06-14 03:20:53 & 60475 & 1364     & 2.2          & 1.6    \\
00032671105 & \swift-XRT (PC)& 2024-06-17 22:33:56 & 2024-06-17 22:55:53 & 60478 & 1311     & 2.1          & 1.5    \\
00032671106 & \swift-XRT (PC)& 2024-06-20 19:51:35 & 2024-06-20 20:12:54 & 60481 & 1274     & 2.0          & 1.4    \\
00032671107 & \swift-XRT (PC)& 2024-06-23 23:36:18 & 2024-06-23 23:56:52 & 60484 & 1229     & 1.8          & 1.5    \\
00032671108 & \swift-XRT (PC)& 2024-06-26 05:19:07 & 2024-06-26 05:38:53 & 60487 & 1181     & 1.7          & 1.5    \\
00032671109 & \swift-XRT (PC)& 2024-06-29 18:17:45 & 2024-06-29 18:40:54 & 60490 & 1381     & 1.7          & 1.4    \\
00032671110 & \swift-XRT (PC)& 2024-07-02 00:03:15 & 2024-07-02 06:35:54 & 60493 & 1518     & 1.6          & 1.6    \\
00032671111 & \swift-XRT (PC)& 2024-07-05 14:43:15 & 2024-07-05 14:58:53 & 60495 & 934      & 1.5          & 1.7    \\
\vspace{0.05cm} \\
91001317002 & \nustar-FPMA   & 2024-04-12 20:41:09 & 2024-04-13 06:36:09 & 60412 & 18919    & 12.4         & -      \\
            & \nustar-FPMB   & 2024-04-12 20:41:09 & 2024-04-13 06:36:09 & 60412 & 18787    & 11.5         & -    \\ 
\vspace{0.05cm} \\
(Archival)  &                &                     &       &          &            \\
80001002002 & \nustar-FPMA   & 2014-01-22 20:16:07 & 2014-01-23 11:36:07 & 56679 & 27754    & 16.3         & -      \\
            & \nustar-FPMB   & 2014-01-22 20:16:07 & 2014-01-23 11:36:07 & 56679 & 27738    & 15.7         & -     \\
80001002004 & \nustar-FPMA   & 2014-01-24 23:56:07 & 2014-01-25 18:31:07 & 56681 & 33237    & 18.5         & -      \\
            & \nustar-FPMB   & 2014-01-24 23:56:07 & 2014-01-25 18:31:07 & 56681 & 33321    & 17.0         & -     \\
            \hline
\end{tabular}

\begin{tablenotes}
    \item[a] Hardness ratio for \swift-XRT: counts ratio of (2-10 keV)/(0.3-2 keV).
\end{tablenotes}
\end{threeparttable}
\end{table*}

\begin{figure*}
    \includegraphics[width=2\columnwidth]{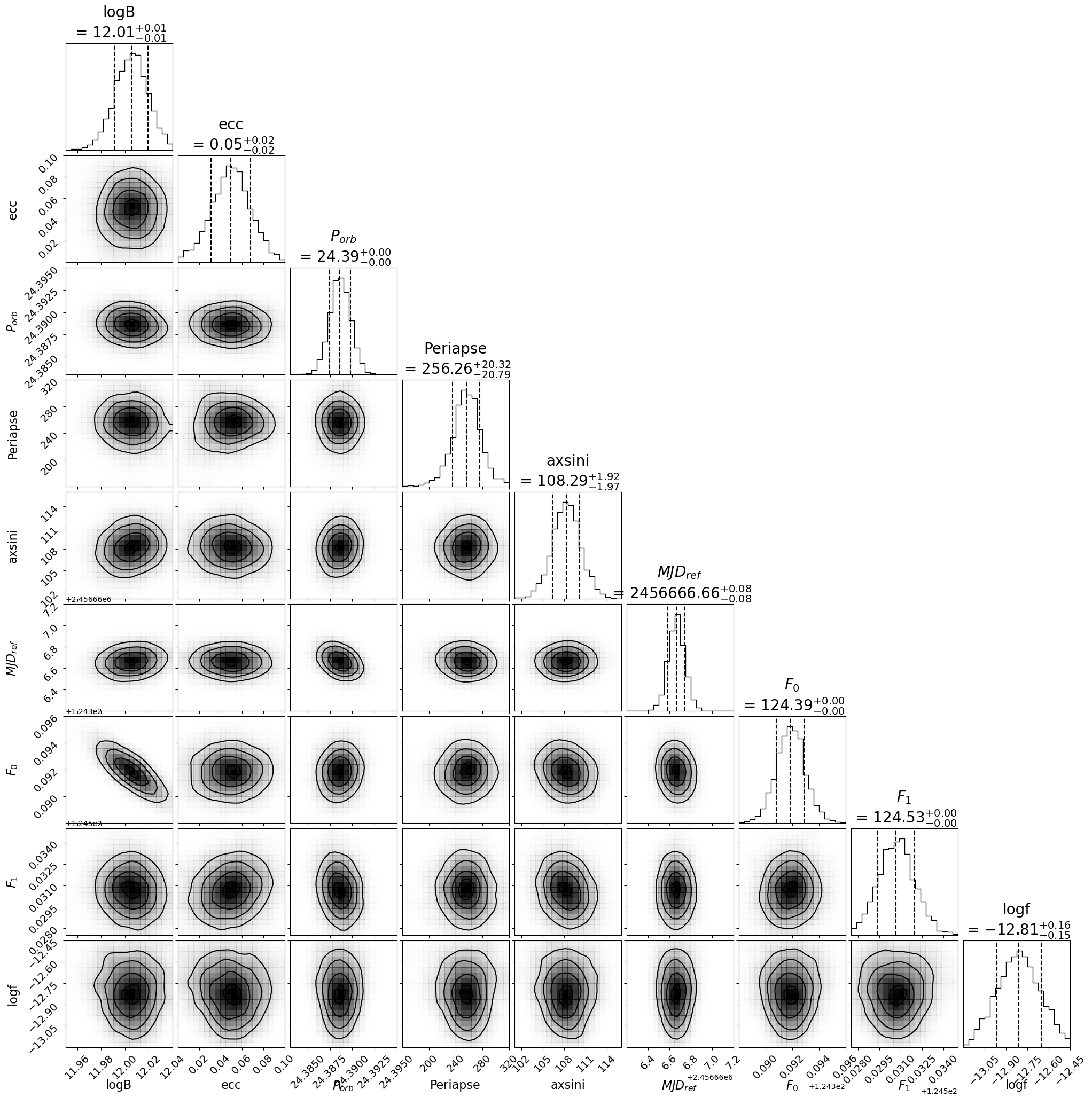}
    \caption{Corner plot of the torque and orbital modelling for the joint fit on the 2014 and 2024 data.}
    \label{fig:time_gbm}
\end{figure*}

\begin{table}
\caption{Joint fit results of additional models}
\label{tab:fit_simul2}
\begin{threeparttable}
\begin{tabular}{llr}
\hline
\multicolumn{1}{c}{Component} & \multicolumn{1}{c}{Parameter} & \multicolumn{1}{c}{Value}            \\ \hline
\multicolumn{3}{l}{\texttt{const*tbabs(nthcomp*gabs+gauss)}}                                                   \\
\texttt{constant} $^{\mathrm{a}}$ & $C_{\mathrm{WT}}$ & $0.888_{-0.115}^{+0.124}$ \\
 & $C_{\mathrm{FPMA}}$ & $1.298_{-0.112}^{+0.128}$ \\
 & $C_{\mathrm{FPMB}}$ & $1.290_{-0.114}^{+0.182}$ \\
\texttt{tbabs} & $N_\mathrm{H}$ & $0.216_{-0.027}^{+0.066}$ \\
\texttt{nthcomp} & Gamma & $1.365_{-0.009}^{+0.008}$ \\
 & $kT_{\mathrm{e}}$ (keV) & $5.01_{-0.20}^{+0.17}$ \\
 & $kT_{\mathrm{bb}}$ (keV) & $0.041_{-0.041}^{+0.080}$ \\
 & norm & $0.026_{-0.003}^{+0.003}$ \\
\texttt{gabs} & $E_{\mathrm{CRSF}}$ (keV) & $32.6_{-0.9}^{+0.9}$ \\
 & $\sigma_{\mathrm{CRSF}}$ (keV) & $8.52_{-0.86}^{+0.80}$ \\
 & Strength & $23.2_{-5.3}^{+5.7}$ \\
\texttt{gauss} & $E_{\mathrm{Fe}}$ (keV) & $6.14_{-0.31}^{+0.19}$ \\
 & $\sigma_{\mathrm{Fe}}$ (keV) & $1.26_{-0.31}^{+0.41}$ \\
 & norm & $0.00082_{-0.00024}^{+0.00029}$ \\
total & cstat/dof & 2734.23/2790 \\
\multicolumn{3}{l}{\texttt{const*tbabs(cutoffpl*gabs+gauss+bbody)}}                                        \\
\texttt{constant} $^{\mathrm{a}}$ & $C_{\mathrm{WT}}$ & $0.894_{-0.112}^{+0.125}$ \\
 & $C_{\mathrm{FPMA}}$ & $1.194_{-0.107}^{+0.120}$ \\
 & $C_{\mathrm{FPMB}}$ & $1.187_{-0.106}^{+0.119}$ \\
\texttt{tbabs} & $N_\mathrm{H}$ & $0.072_{-0.053}^{+0.065}$ \\
\texttt{cutoffPL} & PhoIndex & $0.789_{-0.118}^{+0.130}$ \\
 & HighECut (keV) & $10.1_{-0.8}^{+1.0}$ \\
 & norm & $0.020_{-0.002}^{+0.003}$ \\
\texttt{gabs} & $E_{\mathrm{CRSF}}$ (keV) & $31.0_{-0.6}^{+0.6}$ \\
 & $\sigma_{\mathrm{CRSF}}$ (keV) & $5.22_{-0.47}^{+0.47}$ \\
 & Strength & $24.9_{-4.5}^{+5.7}$ \\
\texttt{gauss} & $E_{\mathrm{Fe}}$ (keV) & $6.49_{-0.13}^{+0.10}$ \\
 & $\sigma_{\mathrm{Fe}}$ (keV) & $0.286_{-0.111}^{+0.198}$ \\
 & norm & $0.000168_{-0.000048}^{+0.000071}$ \\
\texttt{bb} & $kT$ (keV) & $3.97_{-0.13}^{+0.16}$ \\
 & norm & $0.00326_{-0.00040}^{+0.00042}$ \\
total & cstat/dof & 2696.49/2789 \\ \hline

\end{tabular}
\begin{tablenotes}
    \item[a] cross-normalization constants 
\end{tablenotes}

\end{threeparttable}
\end{table}

\begin{table}
\centering
\caption{Spectral Fits to \nustar Observation}
\label{tab:fit_nustar}
\begin{threeparttable}
\begin{tabular}{llr}
\hline
\multicolumn{1}{c}{Component} & \multicolumn{1}{c}{Parameter} & \multicolumn{1}{c}{Value}            \\ \hline
\multicolumn{3}{l}{\texttt{const*tbabs*cflux(powerlaw*fdcut)}}                                                   \\
\texttt{constant}             & $C_{\mathrm{FPMB}}\ ^{\mathrm{a}}$                          & $0.994_{-0.005}^{+0.005}$    \\
\texttt{cflux}                & $\log_{10}$ Flux$^{\mathrm{b}}$ (\uergcms) & $-9.15_{-0.002}^{+0.002}$   \\
\texttt{powerlaw}             & PhoIndex                         & $0.802_{-0.021}^{+0.020}$      \\
\texttt{fdcut}                & $E_{\mathrm{cutoff}}$ (keV)                  & $15.1_{-0.41}^{+0.39}$         \\
                              & $E_{\mathrm{fold}}$ (keV)                    & $5.17_{-0.10}^{+0.10}$       \\
total                         & $\chi^2$/dof                         & $1536.20/1151$                         \\
\multicolumn{3}{l}{\texttt{const*tbabs*cflux(powerlaw*fdcut*gabs)}}                                              \\
\texttt{constant}             & $C_{\mathrm{FPMB}}\ ^{\mathrm{a}}$                          & $0.994_{-0.005}^{+0.005}$    \\
\texttt{cflux}                & $\log_{10}$ Flux$^{\mathrm{b}}$  & $-9.14_{-0.002}^{+0.002}$    \\
\texttt{powerlaw}             & PhoIndex                         & $0.698_{-0.038}^{+0.037}$      \\
\texttt{fdcut}                & $E_{\mathrm{cutoff}}$ (keV)                  & $10.8_{-1.46}^{+1.39}$           \\
                              & $E_{\mathrm{fold}}$ (keV)                    & $7.39_{-0.30}^{+0.31}$         \\
\texttt{gabs}                 & $E_{\mathrm{CRSF}}$ (keV)                    & $31.6_{-0.63}^{+0.68}$         \\
                              & $\sigma_{\mathrm{CRSF}}$ (keV)                & $6.26_{-0.54}^{+0.60}$          \\
                              & Strength                         & $13.4_{-2.30}^{+2.67}$            \\
total                         & $\chi^2$/dof                         & $1343.19/1148$                         \\
\multicolumn{3}{l}{\texttt{const*tbabs*cflux(powerlaw*fdcut*gabs+gauss)}}                                        \\
\texttt{constant}             & $C_{\mathrm{FPMB}}\ ^{\mathrm{a}}$                          & $0.993_{-0.005}^{+0.005}$      \\
\texttt{cflux}                & $\log_{10}$ Flux$^{\mathrm{b}}$ (\uergcms) & $-9.14_{-0.002}^{+0.002}$    \\
\texttt{powerlaw}             & PhoIndex                         & $0.761_{-0.042}^{+0.045}$      \\
\texttt{fdcut}                & $E_{\mathrm{cutoff}}$ (keV)                  & $13.3_{-1.68}^{+2.07}$           \\
                              & $E_{\mathrm{fold}}$ (keV)                    & $7.30_{-0.30}^{+0.31}$         \\
\texttt{gabs}                 & $E_{\mathrm{CRSF}}$ (keV)                    & $31.7_{-0.67}^{+0.72}$         \\
                              & $\sigma_{\mathrm{CRSF}}$ (keV)                & $6.81_{-0.65}^{+0.79}$         \\
                              & Strength                         & $15.4_{-2.90}^{+3.82}$           \\
\texttt{gauss}                & $E_{\mathrm{Fe}}$ (keV)                      & $6.51_{-0.10}^{+0.09}$        \\
                              & $\sigma_{\mathrm{Fe}}$ (keV)                  & $0.28_{-0.10}^{+0.14}$        \\
                              & norm\ $^{\mathrm{c}}$                             & $0.0082_{-0.0022}^{+0.0028}$ \\
total                         & $\chi^2$/dof                         & $1289.75/1145$                         \\
\multicolumn{3}{l}{\texttt{const*tbabs*cflux(nthcomp*gabs+gauss)}}                                               \\
\texttt{constant}             & $C_{\mathrm{FPMB}}\ ^{\mathrm{a}}$                          & $0.994_{-0.005}^{+0.005}$    \\
\texttt{cflux}                & $\log_{10}$ Flux$^{\mathrm{b}}$ (\uergcms) & $-9.14_{-0.002}^{+0.002}$    \\
\texttt{nthcomp}              & Gamma                            & $1.37_{-0.008}^{+0.007}$     \\
                              & $kT_{\mathrm{e}}$ (keV)                      & $4.88_{-0.18}^{+0.14}$         \\
                              & $kT_{\mathrm{bb}}$ (keV)                     & $0.141_{-0.141}^{+0.509}$        \\
\texttt{gabs}                 & $E_{\mathrm{CRSF}}$ (keV)                    & $32.0_{-0.82}^{+0.83}$         \\
                              & $\sigma_{\mathrm{CRSF}}$ (keV)                & $8.00_{-0.72}^{+0.54}$         \\
                              & Strength                         & $20.1_{-4.61}^{+4.85}$           \\
\texttt{gauss}                & $E_{\mathrm{Fe}}$ (keV)                      & $6.16_{-0.23}^{+0.16}$         \\
                              & $\sigma_{\mathrm{Fe}}$ (keV)                  & $1.20_{-0.28}^{+0.34}$         \\
                              & norm\ $^{\mathrm{c}}$                           & $0.0304_{-0.0085}^{+0.0306}$  \\
total                         & $\chi^2$/dof                         & $1305.46/1145$                         \\
\multicolumn{3}{l}{\texttt{const*tbabs*cflux(cutoffpl*gabs+gauss+bbody)}}                                        \\
\texttt{constant}             & $C_{\mathrm{FPMB}}\ ^{\mathrm{a}}$                          & $0.994_{-0.005}^{+0.005}$    \\
\texttt{cflux}                & $\log_{10}$ Flux$^{\mathrm{b}}$ (\uergcms) & $-9.14_{-0.002}^{+0.003}$    \\
\texttt{cutoffPL}             & PhoIndex                         & $1.16_{-0.34}^{+0.30}$         \\
                              & HighECut (keV)                   & $12.9_{-3.09}^{+4.20}$           \\
\texttt{gabs}                 & $E_{\mathrm{CRSF}}$ (keV)                    & $30.1_{-0.58}^{+0.75}$         \\
                              & $\sigma_{\mathrm{CRSF}}$ (keV)                & $4.48_{-0.58}^{+0.68}$         \\
                              & Strength                         & $22.5_{-4.00}^{+5.12}$           \\
\texttt{gauss}                & $E_{\mathrm{Fe}}$ (keV)                      & $6.21_{-0.26}^{+0.34}$         \\
                              & $\sigma_{\mathrm{Fe}}$ (keV)                  & $0.70_{-0.48}^{+0.34}$        \\
                              & norm\ $^{\mathrm{c}}$                             & $0.0142_{-0.0073}^{+0.0069}$ \\
\texttt{bb}                   & $kT$ (keV)                         & $3.75_{-0.09}^{+0.19}$        \\
                              & norm\ $^{\mathrm{c}}$                             & $0.151_{-0.020}^{+0.017}$      \\
total                         & $\chi^2$/dof                         & $1287.12/1144$                         \\ \hline

\end{tabular}
\begin{tablenotes}
    \item[a] cross-normalization constant that is free for FPMB data but frozen to unity for FPMA data.
    \item[b] 3-79 keV flux.
    \item[c] Due to the use of \texttt{cflux} the normalization of another model is related to the normalization of \texttt{powerlaw} or \texttt{nthcomp} or \texttt{cutoffPL}, which is frozen to 1. 
\end{tablenotes}

\end{threeparttable}
\end{table}


\bsp	
\label{lastpage}
\end{document}